\documentclass[letterpaper,12pt,journal,onecolumn]{IEEEtran}

\usepackage{graphicx}
\usepackage{psfig}
\usepackage{amsfonts}
\usepackage{amssymb}
\usepackage{amsmath}

\newtheorem{theorem}{Theorem}[section]
     \newtheorem{lemma}[theorem]{Lemma}
     \newtheorem{proposition}[theorem]{Proposition}

     \newenvironment{example}[1][Example]{\begin{trivlist}
     \item[\hskip \labelsep {\bfseries #1}]}{\end{trivlist}}
     \newenvironment{remark}[1][Remark]{\begin{trivlist}
     \item[\hskip \labelsep {\bfseries #1}]}{\end{trivlist}}

     \newcommand{\qed}{\nobreak \ifvmode \relax \else
           \ifdim\lastskip<1.5em \hskip-\lastskip
           \hskip1.5em plus0em minus0.5em \fi \nobreak
           \vrule height0.75em width0.5em depth0.25em\fi}

\newcounter{proof_steps}

\newcommand{\comment}[1]{}

\begin{document}

\title{Low-Density Parity-Check Code with Fast Decoding Speed }

\author{Xudong Ma \IEEEmembership{student member IEEE} and
En-hui Yang \IEEEmembership{Senior member IEEE}
\thanks{The authors are with the Multimedia Communications Lab,
Electrical and Computer Engineering Department, University of
Waterloo, Waterloo, Ontario, Canada, N2L 3G1. E-mails:
\{x3ma,ehyang\}@bbcr.uwaterloo.ca} \thanks{The material in this
paper was presented in part at the IEEE International Symposium on
Information Theory, Chicago, June 27 - July 2, 2004.}}


\maketitle



\begin{abstract}

Low-Density Parity-Check (LDPC) codes received much attention
recently due to their capacity-approaching performance. The
iterative message-passing algorithm is a widely adopted decoding
algorithm for LDPC codes~\cite{Kschischang01}. An important design
issue for LDPC codes is designing codes with fast decoding speed
while maintaining capacity-approaching performance. In another
words, it is desirable that the code can be successfully decoded
in few number of decoding iterations, at the same time, achieves a
significant portion of the channel capacity. Despite of its
importance, this design issue received little attention so far. In
this paper, we address this design issue for the case of binary
erasure channel.

We prove that density-efficient capacity-approaching LDPC codes
satisfy a so called ``flatness condition''. We show an asymptotic
approximation to the number of decoding iterations. Based on these
facts, we propose an approximated optimization approach to finding
the codes with good decoding speed. We further show that the
optimal codes in the sense of decoding speed are
``right-concentrated''. That is, the degrees of check nodes
concentrate around the average right degree.

\end{abstract}


\section{Introduction}
\label{section_intro}

Low-Density Parity-Check (LDPC) codes are generally decoded by the
iterative message-passing algorithm \cite{Kschischang01}. An
important design issue is finding the codes with fast decoding
speed while maintaining good capacity-approaching performance.
That is, the bit error rate approaches zero with few decoding
iterations while a significant fraction of channel capacity is
achieved. Such codes are desirable because they have less decoding
computational complexity and delay.

Despite of its importance, this design issue received little
attention so far. In this paper, we address this design issue for
the case of Binary Erasure Channel (BEC). BEC is widely adopted as
a practical channel model for packet network communications. In
addition, previous research has shown that the insights gained
from the case of BEC can generally be carried over to the cases of
many other channels.

``Density Evolution'' (DE) is an asymptotic analysis method for
LDPC code performance under the message-passing decoding
\cite{Richardson01}. DE iteratively calculates the probability
distributions of messages for the case where the codeword length
is infinity. For a code with sufficiently long codeword length,
the corresponding distributions of messages are close to these of
the infinitely long codes. Hence, the code performances can be
approximately determined, such as, bit error rates or message
erasure probabilities in the case of BEC.

Given that the message erasure probabilities can be approximately
calculated, a brutal force approach to finding LDPC codes with
good decoding speed is solving a constraint optimization problem.
The optimization variables are the code parameters. The objective
function is the number of decoding iterations, which can be
approximated calculated by DE. And the constraints includes the
fixed code rate, the condition ensuring that the code can be
successfully decoded, and the valid ranges of the code parameters.

Although the above brutal force approach can yield certain codes
with good decoding speed, it is not satisfactory due to the
following reasons. First, the constraint optimization problem does
not have nice numerical properties. The objective function is
discrete and indifferentiable. Almost all optimization algorithms
have convergence problems. It is numerically difficult to find the
optimal solutions.  Second, the approach does not provide a
closed-form relation between the number of iterations and the code
parameters. Therefore, we can not gain any insight into the
problem.

We propose an alternative and tractable approach in this paper. We
prove that ``density-efficient'' and ``capacity-approaching'' LDPC
codes satisfy a so called ``flatness condition''. By
``capacity-approaching'', we mean that the code rate is close to
the channel capacity. By ``density-efficient'', we mean that the
density of the parity-check matrix is low. In this paper, we only
consider codes with efficiently low parity-check matrix density
and low maximal left and right degrees. The codes with high
parity-check matrix density or high maximal degrees are not
practical in implementations.

The flatness-condition simplify our discussion on decoding speed.
Based on that, we present an asymptotic approximation to the
number of decoding iterations. The asymptotic approximation yields
an approximated optimization approach to designing codes with good
decoding speed. Instead of minimizing the number of decoding
iterations directly, the approximated optimization approach
minimizes the asymptotic approximation. Numerical results confirm
that the number of decoding iterations and its asymptotic
approximation are consistent. The approximated optimization
approach also has better numerical properties. The convergence
problem in the brutal force approach is avoided.

The asymptotic approximation provides a closed-form relation
between the number of decoding iterations and the code parameters.
Hence, it provides useful insights into the design problem. One
part of the discussion in this paper on optimal degree
distributions in the sense of decoding speed is based on this
closed-form relation. We also anticipate that this closed-form
relation will facilitate further discussion on decoding speed in
the future.

We also discuss the conditions for the optimal degree
distributions in the sense of decoding speed. We show that the
optimal codes are ``right-concentrated''. That is, the degrees of
check nodes concentrate around the average right degree. In
previous research, several such degree distributions are
numerically found to have nice performance

The remainder of this paper is organized as follows. In Section
\ref{section_pre}, we review LDPC codes, the message-passing
decoding, ``density evolution'', and the density
capacity-approaching tradeoff. Readers who are familiar with these
materials, can skip this section. In Section \ref{section_approx},
we discuss the flatness condition and asymptotic approximation to
the number of decoding iterations. In Section
\ref{section_optim_algo}, we discuss the proposed approximated
optimization approach for finding codes with good decoding speed.
In Section \ref{optimal_degree}, we discuss the conditions for
optimal degree distributions in the sense of decoding speed. In
Section \ref{section_numerical}, we show several numerical
examples. In Section \ref{section_conclusion}, we present our
conclusions.

\section{Preliminary}
\label{section_pre}

\subsection{Binary Erasure Channel}

\begin{figure}[hbt]
\begin{center}
 \includegraphics[width=2in]{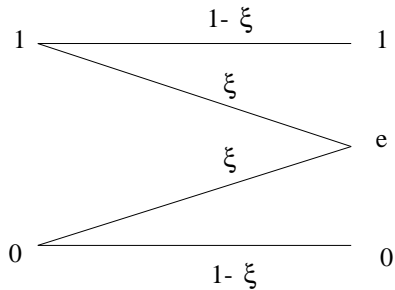}
\end{center}
\caption{The Binary Erasure Channel} \label{fig_bec}
\end{figure}

A BEC is shown in Fig. \ref{fig_bec}. The channel takes binary
inputs and outputs $0$, $1$, or $e$, where $e$ stands for an
erasure. The transmitted binary signal is received correctly with
probability $1-\xi$. Otherwise, the channel outputs an erasure.
The probability $\xi$ is called the channel parameter. The channel
capacity $C=1-\xi$.

\subsection{LDPC code}

\begin{figure}[hbt]
\begin{center}
 \includegraphics[width=2in]{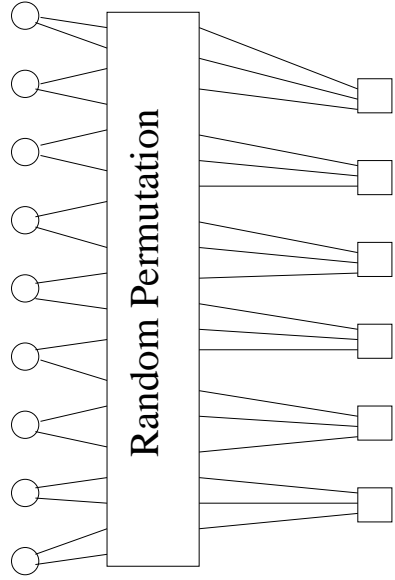}
 \end{center}
 \caption{The Tanner graph representation of Low-Density Parity-Check codes}
 \label{tanner_graph}
\end{figure}

LDPC codes are linear block codes with sparse parity-check matrix.
The codes can be represented by Tanner graphs. A Tanner graph is a
bipartite graph. One of its partition consists of variable nodes;
whereas the other partition consists of check nodes. Each variable
node represents one codeword bit; while each check node represent
one linear check. A Tanner graph is shown in Fig.
\ref{tanner_graph}. The variable nodes are drawn as circles;
whereas the check nodes are drawn as squares.

In this paper, we consider randomly generated LDPC codes. The
Tanner graph is generated according to the three code parameters:
the left degree distribution $\lambda(x)$, the right degree
distribution $\rho(x)$, and the codeword length $N$. The left and
right degree distributions are polynomials:
\begin{equation}
\lambda(x)=\sum_{i=2}^{d_v}\lambda_ix^{i-1}
\end{equation}
~
\begin{equation}
\rho(x)=\sum_{j=2}^{d_c}\rho_jx^{j-1}
\end{equation}
where $\lambda_i$ is the fraction of edges connected to variable
nodes with degree $i$; while $\rho_j$ is the fraction of edges
connected to check nodes with degree $j$. The Tanner graph is
generated by first growing edges from variable and check nodes
according to the degree distributions. The edges from variable
nodes and check nodes are then uniformly randomly connected. If
the codeword length is sufficiently long, approximately the code
rate
\begin{equation}
R=1-\frac{\int_{0}^{1}\rho(x)dx}{\int_{0}^{1}\lambda(x)dx}
\end{equation}

\subsection{Message-passing Decoding}

The message-passing algorithm \cite{Kschischang01} is an iterative
decoding algorithm. The algorithm computes likelihood functions
for all codeword bits iteratively. The final decoding decisions
are hard threshold decisions on the likelihood functions.

In the case of BEC, the likelihood function has a finite alphabet.
The message-passing algorithm becomes simple. During each
iteration, the algorithm find check nodes with only one
neighboring variable node being still erasure. The algorithm then
corrects these erasures according to the linear constrains.

\subsection{Density Evolution}

Density Evolution \cite{Richardson01} calculates the distributions
of messages for the codes with infinitely long codeword length. In
the case of BEC, the message erasure probability $P_e^{(l)}$ after
the $l$-th iteration can be approximately calculated as follows:
\begin{equation}
P_e^{(l)}=\xi\lambda(1-\rho(1-P_e^{(l-1)}))
\end{equation}
Asymptotically with the codeword length, the code can be
successfully decoded with high probability if and only if
\begin{equation}
\xi\lambda(1-\rho(1-x))<x\,\,\,\mbox{ for any }x\in(0,\xi]
\label{eqn_asym_sucess}
\end{equation}

\subsection{Density Capacity-approaching Tradeoff}
\label{subsec_dc_tradeoff}

It is shown that LDPC codes with bounded degrees can not achieve
the channel capacity. There exists a tradeoff between the
parity-check matrix density and the achievable rate
\cite{Gallager} \cite{Sason03}. In the case of BEC, Shokrollahi
\cite{Shokrollahi99} shows the following bound for achievable
rate:
\begin{equation}
(1-\xi)^a\leq \frac{\Delta R}{\xi+\Delta R}
\end{equation}
where, $a$ is the average right degree,
\begin{equation}
1/a=\int_{0}^{1}\rho(x)dx
\end{equation}
and $\Delta R$ is the gap between the achievable rate to the
channel capacity. This lower bound of the gap to the capacity
decreases exponentially as the average degree increases. In the
same paper, Shokrollahi show that this bound is tight. That is,
there exist codes with an exponentially decreasing gap to the
capacity with respect to linearly growing average degrees.

\section{Flatness Condition and Asymptotic Approximation to the Number of Iterations}
\label{section_approx}

In this section, we prove that density-efficient
capacity-approaching LDPC codes satisfy a necessary condition -
the flatness condition. Based on the flatness condition, we
further derive an asymptotic approximation to the number of
decoding iterations.

The propositions and lemmas in Section
\ref{subsec_auxiliary_lemma} are not meaningful by themselves.
They are useful only for proving the main theorems in Section
\ref{sub_sec_main_theorem}. A reader who is not interesting in the
details of the proof can skip the section
\ref{subsec_auxiliary_lemma}.

\subsection{Notation and Definition}

Consider a BEC with channel parameter $\xi$. Consider a LDPC code
with the left degree distribution $\lambda(x)$ and the right
degree distribution $\rho(x)$. Denote the average right degree by
$a$. Define $b=1/a$. Let $R$ denote the code rate. The gap between
the capacity and the code rate $\Delta R= C-R$. We define the
function $B(\Delta R, b,x)$ as follows:
\begin{equation}
B(\Delta R, b,x)=\sqrt{\frac{2 \Delta R}{(\xi+\Delta
R)(1-\xi)\rho(1-x)}}
\end{equation}

We define the {\it decoding convergence time} $T_{\eta}$ to be the
maximal $l$ such that the message erasure probability is greater
than the probability level $\eta$ after $l$ decoding iterations.
For any left and right degree distributions $\lambda(x)$ and
$\rho(x)$, we define
\begin{equation}
F(\lambda,\rho,\eta)\triangleq \int_{\eta}^{\xi}
\frac{1}{x-\xi\lambda(1-\rho(1-x))}dx
\end{equation}

Throughout Section \ref{section_approx}, the derivatives are taken
with respect to $x$. Throughout this paper, we also assume that
the maximal left and right degrees are upper bounded by $k_va$ and
$k_ca$ respectively, where $k_v$ and $k_c$ are constants.

\subsection{Auxiliary Propositions and Lemmas}
\label{subsec_auxiliary_lemma}

 The proofs can be found in the Appendices
\ref{proof_pro1} to \ref{proof_lemma_boundary}.

%
%

\begin{proposition}
\begin{equation}
\rho(1-x)\leq (-1)[\rho(1-x)]'\leq
\frac{k_ca}{1-\xi}\rho(1-x),\,\,\,\mbox{ for any }x\in(0,\xi]
\end{equation}
\label{pro1}
\end{proposition}

\begin{proposition}
\begin{eqnarray}
[\rho(1-x)]''\leq (-1)\frac{k_ca}{1-\xi}[\rho(1-x)]',\,\,\,\mbox{
for any }x\in(0,\xi] \nonumber \
\end{eqnarray}
\label{pro2}
\end{proposition}
\begin{proof}
Similar to that of Proposition \ref{pro1}.

\end{proof}

%
%

\begin{lemma}
\begin{equation}
\int_{0}^{1}\rho(1-x)dx-\int_{0}^{\xi}\left[1-\lambda^{-1}(x/\xi)\right]dx=
\frac{b\Delta R}{\xi+\Delta R} \label{eqn_bound_integral_of_rho}
\end{equation}
\label{lemma_bound_integral_of_rho}
\end{lemma}

%
%

\begin{lemma}
\begin{eqnarray}
\frac{1-\lambda^{-1}(x/\xi)} {\rho(1-x)}\geq 1-\sqrt{k_c}B(\Delta
R,b,x),\,\,\, \mbox{ for any }x\in (0,\xi)
\end{eqnarray}
\label{lemma_rho_lemma_diff_bound}
\end{lemma}

%
%

\begin{lemma}
\begin{equation}
\left|[\lambda(1-\rho(1-x))]''\right|\leq
\frac{k_v^2k_c^2\rho(1-x)^2a^4}{(1-\xi)^2}+
\frac{k_vk_c^2\rho(1-x)a^3}{(1-\xi)^2}
\end{equation}
\label{lemma_bound_recursive_eqn}
\end{lemma}

%
%

\begin{lemma}
\label{lemma_bound_ratio_derivatives} Let $b^5+b^2 < \xi$ and
$b^5<x<\xi-b^2$. Then,
\begin{eqnarray}
 \frac{[1-\lambda^{-1}(x/\xi)]'}{[\rho(1-x)]'} \geq
1+\frac{-k_cb}{2(1-\xi)} -  \sqrt{k_c}B(\Delta R,b,x)(1-\xi)a^2
 \nonumber \\
\end{eqnarray}
\begin{eqnarray}
\frac{[1-\lambda^{-1}(x/\xi)]'}{[\rho(1-x)]'}  \leq 1+
\sqrt{k_c}B(\Delta R,b,x)a^5+
\frac{k_cb^4}{2(1-\xi)}(1+\frac{b^5}{1-\xi})^{k_ca} \nonumber
\\
\end{eqnarray}
\end{lemma}

%
%

\begin{lemma}
\label{lemma_bound_derivatives} Let $x_0\in (0,\xi)$ and $
x_1=\xi\lambda(1-\rho(1-x_0))$. Then,
\begin{equation}
[\xi\lambda(1-\rho(1-x_0))]' \leq
\frac{[\rho(1-x_i)]'}{[1-\lambda^{-1}(x_i/\xi)]'}
\end{equation}
for $i=0,1$.
\end{lemma}

%
%

\begin{lemma}
\label{lemma_bound_two_curve_y}For $x\in (0,\xi)$,
\begin{equation}
x-\xi\lambda(1-\rho(1-x)) \leq \sqrt{ \frac{-2b\Delta
R}{(\xi+\Delta R)[\rho(1-x)]'}}
\end{equation}
\end{lemma}

\comment{
\begin{lemma}
\label{lemma_bound_two_curve_y}Let $x_0\in (0,\xi)$ and
$x_1=\xi\lambda(1-\rho(1-x_0)) $.
\begin{equation}
\Delta x=x_0-x_1\leq \sqrt{ \frac{-2b\Delta R}{(\xi+\Delta
R)[\rho(1-x_0)]'}}
\end{equation}
\end{lemma}
}

%
%

\begin{lemma}
\label{lemma_boundary} If $b<(1-\xi)/k_c$, then
\begin{equation}
\rho(1-\xi+b^2)\leq \frac{2k_c(1-\xi)\Delta R}{(\xi+\Delta
R)(1-\xi-k_cb^2)}
\end{equation}
\end{lemma}

\subsection{Main Theorems}
\label{sub_sec_main_theorem}

%
%

Let us consider a BEC with channel parameter $\xi$. Consider a
sequence of degree distribution pairs:
\begin{equation}
(\lambda_1,\rho_1),\cdots,(\lambda_n,\rho_n),\cdots
\end{equation}
where
\begin{equation}
\lambda_n(x)=\sum_{i}\lambda_{ni}x^{i-1}
\end{equation}
\begin{equation}
\rho_n(x)=\sum_{j}\rho_{nj}x^{j-1}
\end{equation}
are the left and right degree distributions respectively. Each
pair of degree distributions defines a code. Let $a_n$ denote the
average right degree for the $n$-th code. Define $b_n=1/a_n$. Let
$R_n$ denote the rate of the $n$-th code. Let $\Delta R_n$ denote
the gap between the capacity and the code rate for the $n$-th
code, $\Delta R_n=C-R$.

Then we have the following theorem.

\begin{theorem}
If $b_n$ is strictly decreasing with $\lim_{n\rightarrow
\infty}b_n=0$. The gap $\Delta R_n$ is strictly decreasing and
\begin{equation}
\lim_{n\rightarrow \infty}\frac{\Delta R_n}{(b_n)^{15}}\rightarrow
0
\end{equation}
The successfully decoding condition holds:
\begin{equation}
\xi\lambda_n(1-\rho_n(1-x))<x,\,\,\,\mbox{for any }x\in(0,\xi]
\end{equation}
The maximal degrees of $\lambda_n$ and $\rho_n$ are upper bounded
by $k_va_n$ and $k_ca_n$ respectively, where $k_v$ and $k_c$ are
constants. Then as $n\rightarrow \infty$, the derivative of
$\xi\lambda_n(1-\rho_n(1-x))$ with respect to $x$ converges to $1$
uniformly with respect to $x$ in the interval $(0,\xi]$.
\label{theorem_flatness_main}
\end{theorem}
\begin{proof}
The proof is in Appendix \ref{proof_theorem_flatness_main}
\end{proof}

According to this theorem, the function $\xi\lambda(1-\rho(1-x))$
becomes flat as the code rate approaches the channel capacity and
the parity-check matrix density remains efficiently low. Based on
this conclusion, we prove the following theorem, which shows an
asymptotic approximation to decoding convergence time $T_\eta$.

%
%

\begin{theorem}
Let $\eta$ be a fixed probability level, $0<\eta<\xi$. If $b_n$ is
strictly decreasing with $\lim_{n\rightarrow \infty}b_n=0$. The
gap $\Delta R_n$ is strictly decreasing and
\begin{equation}
\lim_{n\rightarrow \infty}\frac{\Delta R_n}{(b_n)^{15}}\rightarrow
0
\end{equation}
The successfully decoding condition holds:
\begin{equation}
\xi\lambda_n(1-\rho_n(1-x))<x,\,\,\,\mbox{for any }x\in(0,\xi]
\end{equation}
The maximal degrees of $\lambda_n$ and $\rho_n$ are upper bounded
by $k_va_n$ and $k_ca_n$ respectively, where $k_v$ and $k_c$ are
constants. Then as $n\rightarrow \infty$, the ratio
\begin{equation}
F(\lambda_n,\rho_n,\eta)/T_\eta
\end{equation}
goes to $1$.
 \label{theorem_asym_approx}
\end{theorem}
\begin{proof}
The proof is in Appendix \ref{proof_theorem_asym_approx}.
\end{proof}

\begin{remark}
In the above theorems, we assume that $\Delta R_n$ decrease
polynomially with respect to $b_n$. According to Section
\ref{subsec_dc_tradeoff}, in the most efficient tradeoffs, $\Delta
R_n$ decreases exponentially with respect to $b_n$. We conclude
that this condition is generally satisfied in practical
applications.
\end{remark}

\section{Approximated Optimization}

\label{section_optim_algo}

A brutal force approach to finding the LDPC code with minimal
decoding convergence time $T_\eta$ and a fixed gap to the capacity
is solving the following constrain optimization problem:
\begin{equation}
\min T_\eta \label{direct_cons_prob_1}
\end{equation}
subject to
\begin{equation}
\xi\lambda(1-\rho(1-x))<x \mbox{   for }x\in(0,\xi]
\label{direct_cons_prob_2}
\end{equation}
\begin{equation}
\frac{\int_{0}^{1}\rho(x)dx}{\int_{0}^{1}\lambda(x)dx}= 1-C+\Delta
R \label{direct_cons_prob_3}
\end{equation}
\begin{equation}
\sum_{i}\lambda_i=1,\,\,\,\sum_{j}\rho_j=1,\,\,\,\lambda_i\geq
0,\mbox{ and }\rho_j\geq 0 \label{direct_cons_prob_4}
\end{equation}
Where $C$ is the channel capacity, and $0<\Delta R<C$ is a fixed
gap to the channel capacity. The condition in Eqn.
\ref{direct_cons_prob_2} is the successful decoding condition. The
condition in Eqn. \ref{direct_cons_prob_3} imposes that the code
rate is $C-\Delta R$. The constraints in Eqn.
\ref{direct_cons_prob_4} come from the probability nature of
degree distributions.

However, the above optimization problem is not tractable. The
objective function $T_\eta$ is not differentiable. This brings in
convergence problems for optimization algorithms. To get around
these difficulties, at this point, we invoke Theorem
\ref{theorem_asym_approx} and replace $T_\eta$ by
$F(\lambda,\rho,\eta)$. This yields the following approximated
optimization problem:
\begin{equation}
\min F(\lambda,\rho,\eta) \label{approx_cons_prob_1}
\end{equation}
subject to
\begin{equation}
\xi\lambda(1-\rho(1-x))<x,\,\,\,\mbox{ for }x\in(0,\xi]
\label{approx_cons_prob_2}
\end{equation}
\begin{equation}
\frac{\int_{0}^{1}\rho(x)dx}{\int_{0}^{1}\lambda(x)dx}= 1-C+\Delta
R \label{approx_cons_prob_3}
\end{equation}
\begin{equation}
\sum_{i}\lambda_i=1,\,\,\,\sum_{j}\rho_j=1,\,\,\,\lambda_i\geq
0,\mbox{ and }\rho_j\geq 0 \label{approx_cons_prob_4}
\end{equation}
The numerical results confirm that the approximated optimization
has nice numerical properties.

\section{Optimal Degree Distribution}
\label{optimal_degree}

In this section, we discuss the conditions for the optimal degree
distributions in the sense of convergence speed. We show that the
optimal codes are right-concentrated.

\subsection{Low Erasure Probability Region Convergence Speed
Analysis}

In practical applications, the probability level $\eta$ is
generally small. The number of decoding iterations while the
erasure probability is in a low erasure probability region may
constitute a large fraction of the decoding convergence time. That
is, the decoding speed in the low erasure probability region
dominates the overall decoding speed.

We show in this section that if the average right degree is fixed,
then the right-concentrated degree distributions have optimal
decoding speed in the low erasure probability region.

\begin{figure}[hbt]
\begin{center}
 \includegraphics[width=5in]{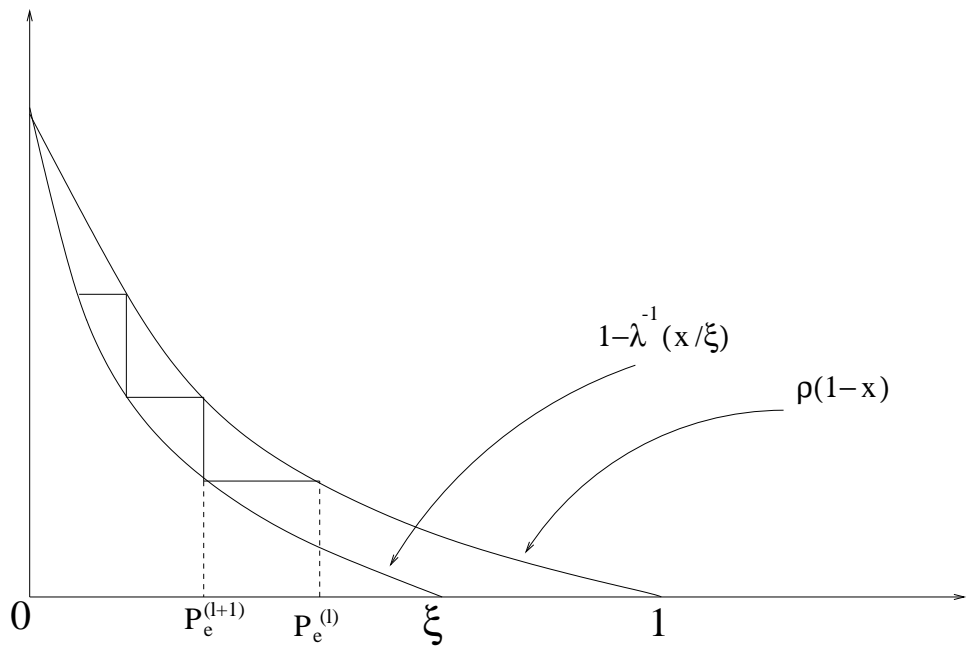}
 \end{center}
 \caption{The iteration process of erasure probabilities}
 \label{fig_iter_process}
\end{figure}

Note that the relation between $P_e^{(l)}$ and $P_e^{(l+1)}$ can
be also written as follows:
\begin{equation}
y=\rho(1-P_e^{(l)})
\end{equation}
\begin{equation}
y=1-\lambda^{-1}(P_e^{(l+1)}/\xi)
\end{equation}
where $y$ is a auxiliary variable. The iterative process of
$P_e^{(l)}$ is illustrated in Fig. \ref{fig_iter_process}. To
increase the decoding speed, we need to move the curve $\rho(1-x)$
upward and the curve $1-\lambda^{-1}(x/\xi)$ to the left. Moving
the curve $\rho(1-x)$ upward is equivalent to making the function
$\rho(1-x)$ larger. Moving the curve $1-\lambda^{-1}(x/\xi)$ to
the left is equivalent to making the function $\lambda(1-x)$
smaller. To have fast decoding speed in the low message erasure
probability region, we need to make $\rho(x)$ large for $x$ near
$1$ and $\lambda(x)$ small for $x$ near $0$.

We need the following auxiliary lemma for proving the main theorem
in this section.

%
%

\begin{lemma}
\label{lemma_basic_degree_distribution1} Let $\gamma(x)$ be a
degree distribution with average degree $a$ and maximal degree
$d\geq 3$. Assume $\gamma_{i}>0$ and $\gamma_{i+2}>0$, for
$1<i<d$. Then another degree distribution $\hat{\gamma}(x)$ with
the same average and maximal degrees can be constructed as
follows:
\begin{align}
\hat{\gamma}(x)= & { \gamma(x) + \beta x^{i}
-\left\{\frac{i}{2(i+1)}\beta
x^{i-1}+\frac{i+2}{2(i+1)}\beta x^{i+1}\right\} } \notag \\
\end{align}
where $\beta$ is a sufficiently small positive real number such
that $\hat{\gamma}(x)$ is well defined. We also have
\begin{equation}
\gamma(x)< \hat{\gamma}(x), \,\,\, \mbox{ for }  \frac{i}{i+2}<x<1
\end{equation}
\begin{equation}
\gamma(x)> \hat{\gamma}(x),\,\,\, \mbox{ for } 0<x< \frac{i}{i+2}
\end{equation}
\end{lemma}
\begin{proof}
The proof is in Appendix
\ref{proof_lemma_basic_degree_distribution1}.
\end{proof}

\comment{
\begin{theorem}
\label{theorem_gamma_min} Let $x$ be an arbitrary real number,
$0<x<1/2$. Then the degree distribution with average degree $a$
and maximal degree $d$ which minimizes the function $\gamma(x)$ is
\begin{equation}
\gamma(x)=\gamma_2x+\gamma_dx^{d-1}
\end{equation}
\end{theorem}
\begin{proof}
The proof is in Appendix \ref{proof_theorem_gamma_min}.
\end{proof}
}

\begin{theorem}
\label{theorem_gamma_max} Let $d$ be an positive integer. Let $x$
be an arbitrary real number, $x>1-\frac{2}{d+1}$. Then the degree
distribution with average degree $a$ and maximal degree $d$ which
maximizes the function $\gamma(x)$ is
\begin{equation}
\gamma(x)=\gamma_ix^{i-1}+\gamma_{i+1}x^{i}
\end{equation}
where $i=\lfloor a \rfloor$ is the largest integer smaller than
$a$.
\end{theorem}
\begin{proof}
The proof is in Appendix \ref{proof_theorem_gamma_max}.
\end{proof}

The above Theorems imply that the degree distributions with fast
decoding speed at the low erasure probability region are
right-concentrated.

\subsection{Asymptotic Approximation Based Analysis}

In this section, we discuss the condition for the optimal degree
distributions in the approximated constraint optimization program
in Eqn. \ref{approx_cons_prob_1}, \ref{approx_cons_prob_2},
\ref{approx_cons_prob_3}, \ref{approx_cons_prob_4}. We have the
following theorem, which shows that the optimal degree
distributions are right-concentrated.

\begin{theorem}
Let $(\lambda^\ast,\rho^\ast)$ be the solution of the constrain
optimization problem in Eqns. \ref{approx_cons_prob_1},
\ref{approx_cons_prob_2}, \ref{approx_cons_prob_3},
\ref{approx_cons_prob_4} with fixed maximal left degree $d_v$ and
maximal right degree $d_c$. Then
\begin{itemize}
\item If $\xi\leq 1-e^{-2/d_c}$, then $\rho_j^\ast$ are non-zero
only for two $j$'s.

\item If $\xi>1-e^{-2/d_c}$, then either
\begin{equation}
\min_{x\in
(1-e^{-2/d_c},\xi)}[x-\xi\lambda^\ast(1-\rho^\ast(1-x))]\rightarrow
0\end{equation} as $\eta\rightarrow 0$ or $\rho_j^\ast$ are
non-zero only for two $j$'s when $\eta$ is sufficiently small.

\end{itemize}
\label{right_concen_main_thm}
\end{theorem}
\begin{proof}
The proof is in Appendix \ref{proof_right_concen_main_thm}.
\end{proof}

\section{Numerical Examples}
\label{section_numerical}

In this section, we show several numerical design examples. We
also compare the codes designed by the proposed approach to one
well-known class of LDPC codes: the Heavy-tail/Poisson codes
\cite{Luby01} \cite{Shokrollahi00}.

\begin{figure}[hbt]
\begin{center}
 \includegraphics[width=5in]{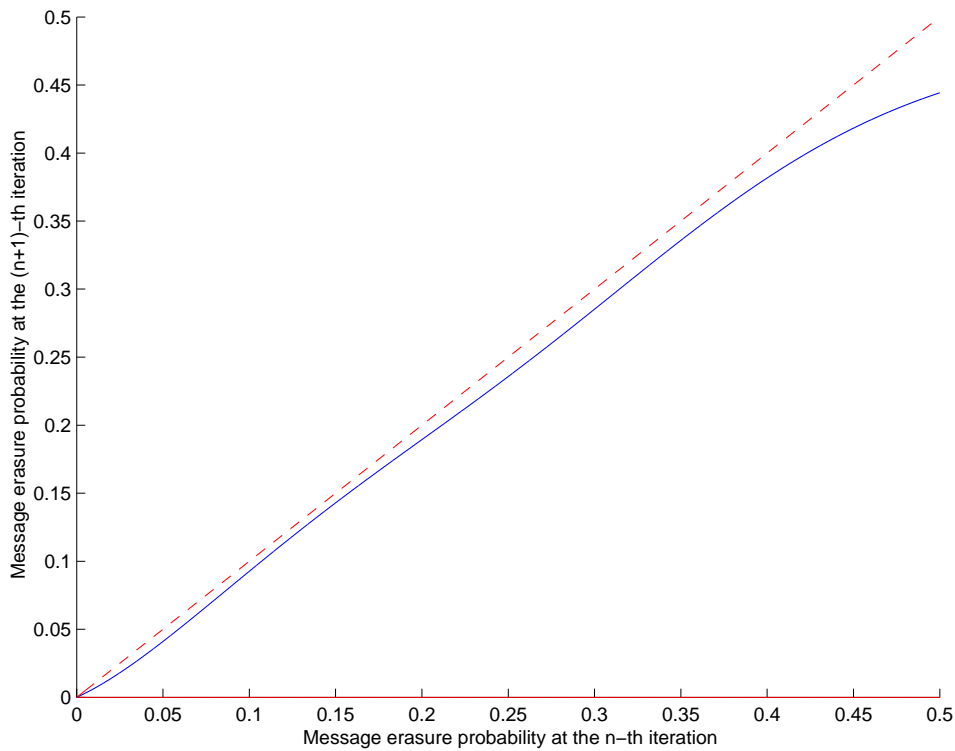}
 \end{center}
 \caption{The function $\xi\lambda(1-\rho(1-x))$ in the Example 1}
 \label{aa1}
\end{figure}

\begin{figure}[hbt]
\begin{center}
 \includegraphics[width=5in]{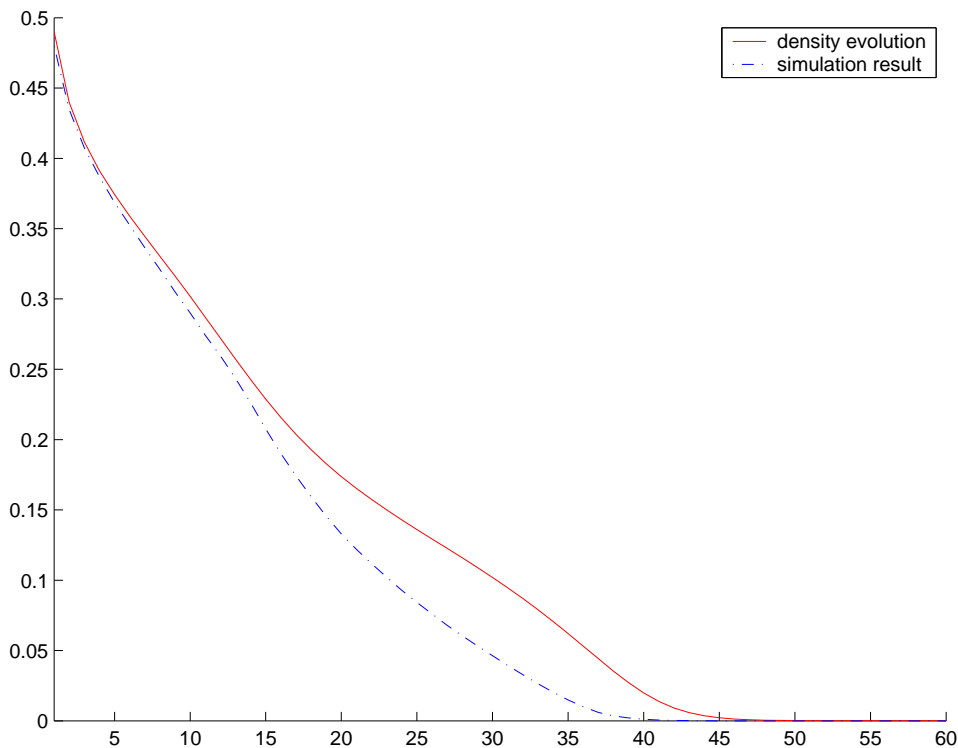}
 \end{center}
 \caption{The message erasure probabilities in the example 1}
 \label{example1b}
\end{figure}

\begin{example}[Example 1:]
In this example, we design a LDPC code for a BEC with parameter
$\xi=0.48$. The code rate is $0.48$. The left and right degree
distributions are as follows:
\begin{equation}
\lambda(x)=0.1863x+0.4143x^2+0.0512x^8+0.3482x^{15}
\end{equation}
\begin{equation}
\rho(x)=0.5330x^6+0.4670x^7
\end{equation}
The function $\xi\lambda(1-\rho(1-x))$ is shown in Fig. \ref{aa1}
as the solid line. The dashed line shows the straight line $y=x$.
For $\eta=10^{-3}$, The decoding convergence time $T_{\eta}=47$ in
the corresponding density evolution. The asymptotic approximation
$F(\lambda,\rho,\eta)=47.9400$.

We construct practical codes according to the designed degree
distributions. The codeword length is $32k$ bits. The simulation
results on message erasure probabilities are shown in Fig.
\ref{example1b}. The message erasure probabilities after different
numbers of iterations are shown as the dash-dot curve. The message
erasure probabilities by density evolution are shown as the solid
curve.
\end{example}

\begin{figure}[hbt]
\begin{center}
 \includegraphics[width=5in]{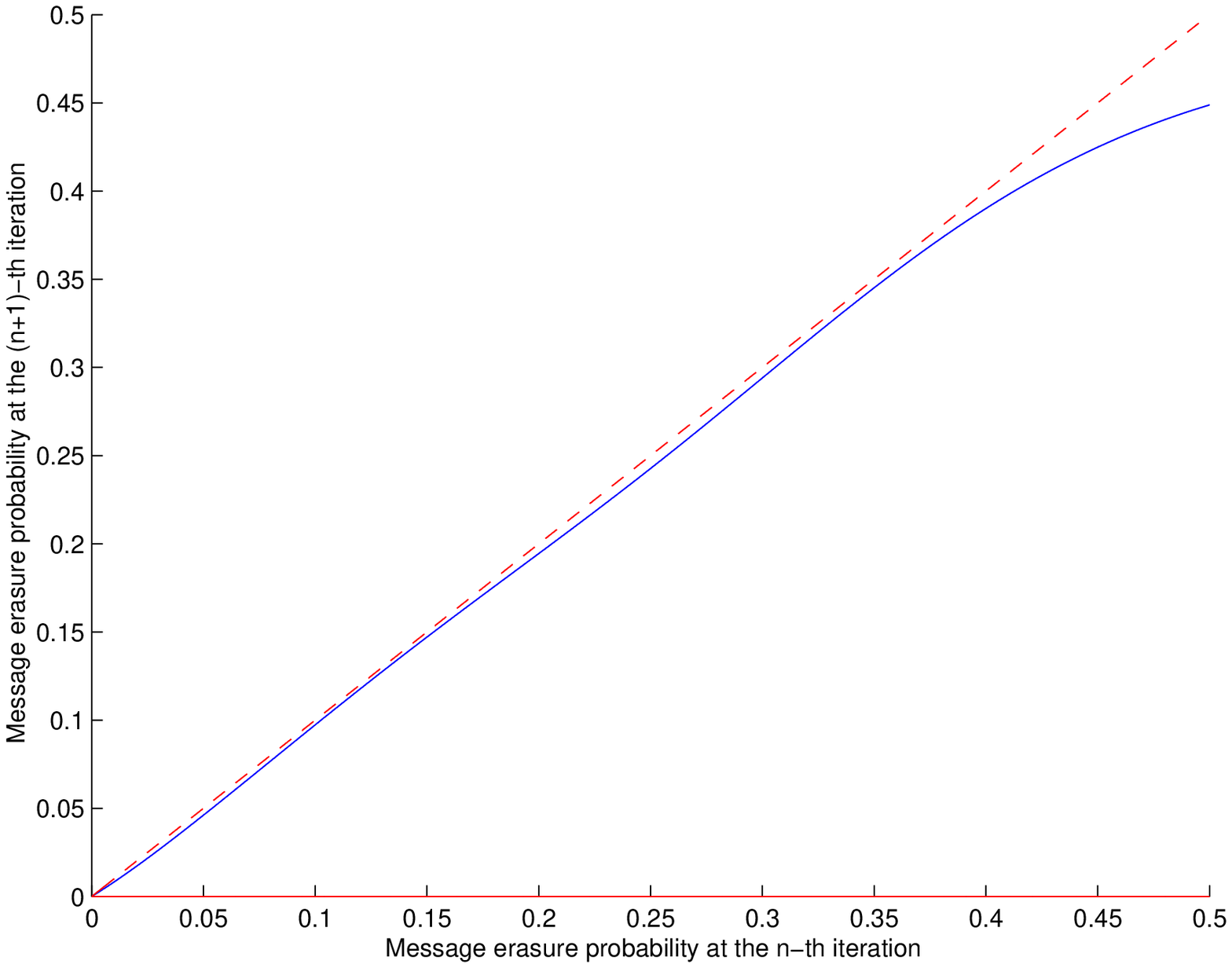}
 \end{center}
 \caption{The function $\xi\lambda(1-\rho(1-x))$ in the Example 2}
 \label{aa2}
\end{figure}

\begin{figure}[hbt]
\begin{center}
 \includegraphics[width=5in]{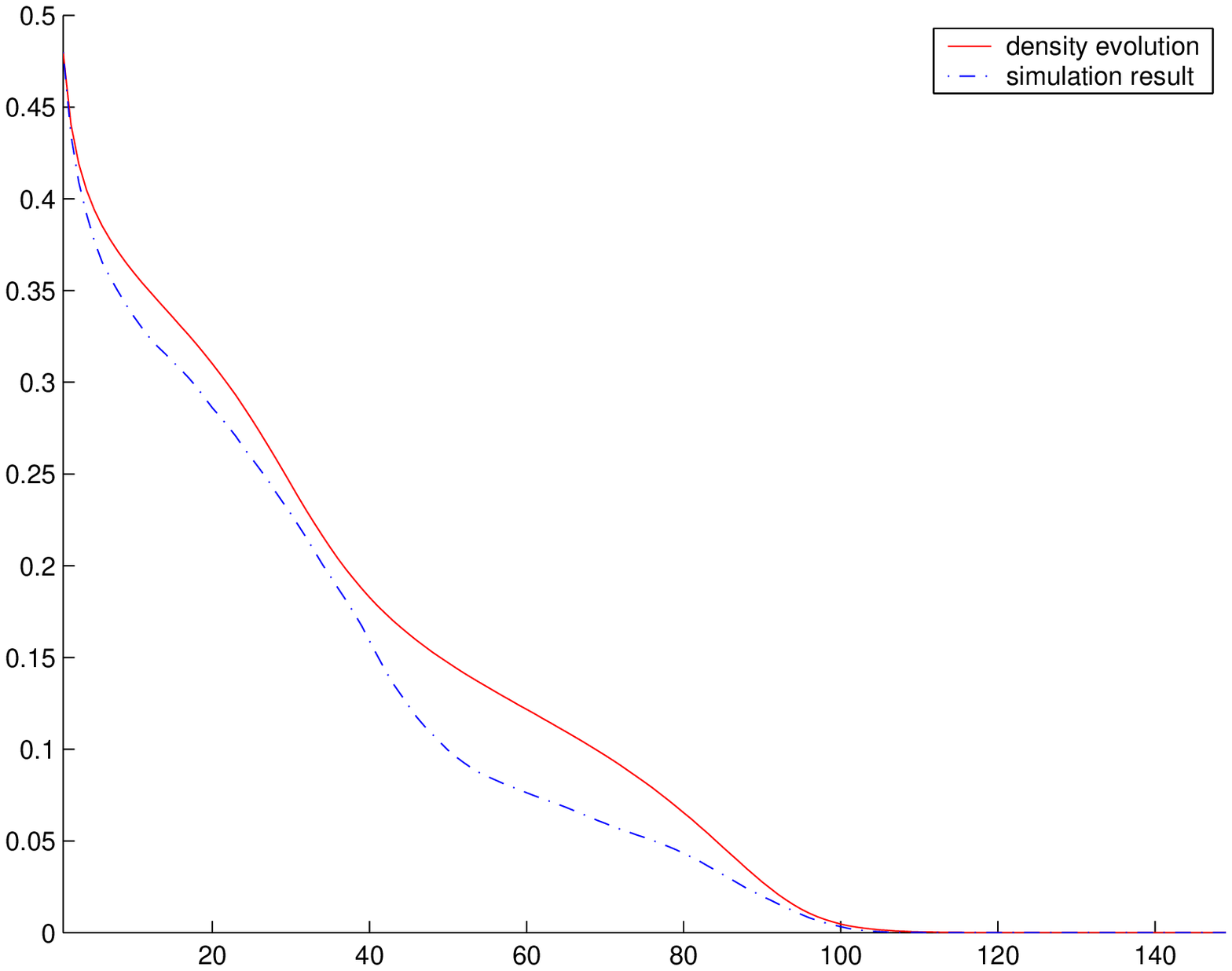}
 \end{center}
 \caption{The message erasure probabilities in the Example 2}
 \label{example2b}
\end{figure}

\begin{example}[Example 2:]
In the second example, we consider the BEC with parameter
$\xi=0.48$. The code rate is $0.5$. We find the following left and
right degree distributions:
\begin{equation}
\lambda(x)=0.2452x+0.2982x^2+0.1112x^5+0.3454x^{15}
\end{equation}
\begin{equation}
\rho(x)=0.3398x^6+0.6602x^7
\end{equation}
The function $\xi\lambda(1-\rho(1-x))$ is shown in Fig. \ref{aa2}
as the solid line. The dash line shows the straight line $y=x$.
For $\eta=10^{-3}$, the decoding convergence time $T_{\eta}=107$
in the corresponding density evolution. The asymptotic
approximation $F(\lambda,\rho,\eta)=108.7363$.

We construct practical codes according to the above degree
distributions. The simulation results on message erasure
probabilities after different numbers of iterations are shown in
Fig. \ref{example2b} as the dash-dot curve. The message erasure
probabilities by density evolution are shown as the solid curve.
The codeword length is $32k$ bits.
\end{example}

\begin{figure}[hbt]
\begin{center}
 \includegraphics[width=5in]{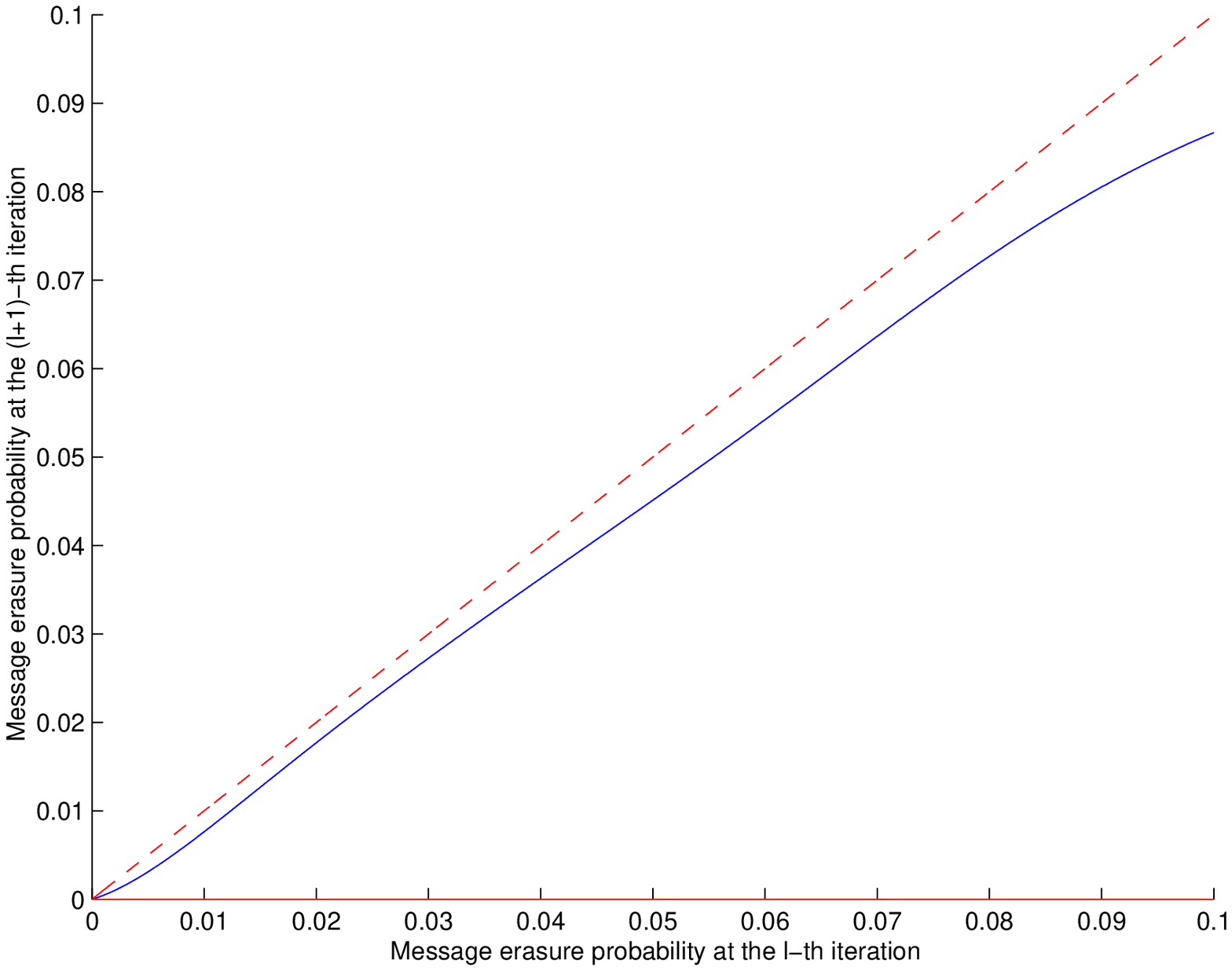}
 \end{center}
 \caption{The function $\xi\lambda(1-\rho(1-x))$ in the Example 3}
 \label{aa3}
\end{figure}

\begin{figure}[hbt]
\begin{center}
 \includegraphics[width=5in]{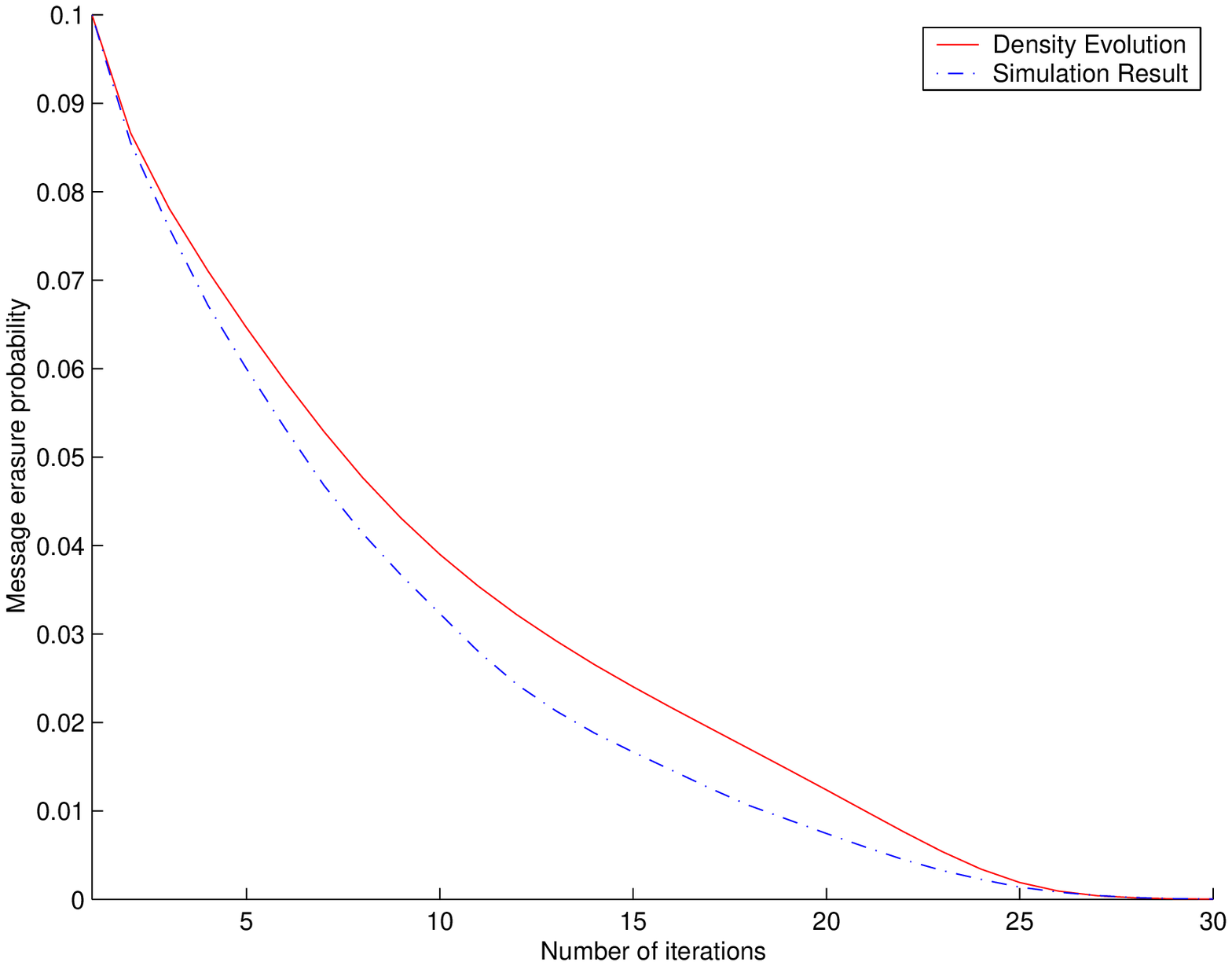}
 \end{center}
 \caption{The message erasure probabilities in the Example 3}
 \label{example3b}
\end{figure}

\begin{example}[Example 3:]
In the third example, we design codes for the BEC with parameter
$\xi=0.1$. The code rate is $0.885$. The left and right degree
distributions are as follows:
\begin{equation}
\lambda(x)=0.0939x+0.3807x^2+0.0443x^9+0.1875x^{10}+0.2937x^{31}
\end{equation}
\begin{equation}
\rho(x)= 0.4725x^{41}+0.5274x^{42}
\end{equation}
The function $\xi\lambda(1-\rho(1-x))$ is shown in Fig. \ref{aa3}
as the solid curve. The dash line shows the straight line $y=x$.
For $\eta=10^{-3}$, the decoding convergence time $T_{\eta}=26$ in
the corresponding density evolution. The asymptotic approximation
$F(\lambda,\rho,\eta)=26.6844$.

The simulation results on the message erasure probabilities after
different numbers of iterations are shown in Fig. \ref{example3b}.
The message erasure probabilities in the density evolution are
shown as the solid curve. The codeword length is $32k$ bits.
\end{example}

\begin{figure}[hbt]
\begin{center}
 \includegraphics[width=5in]{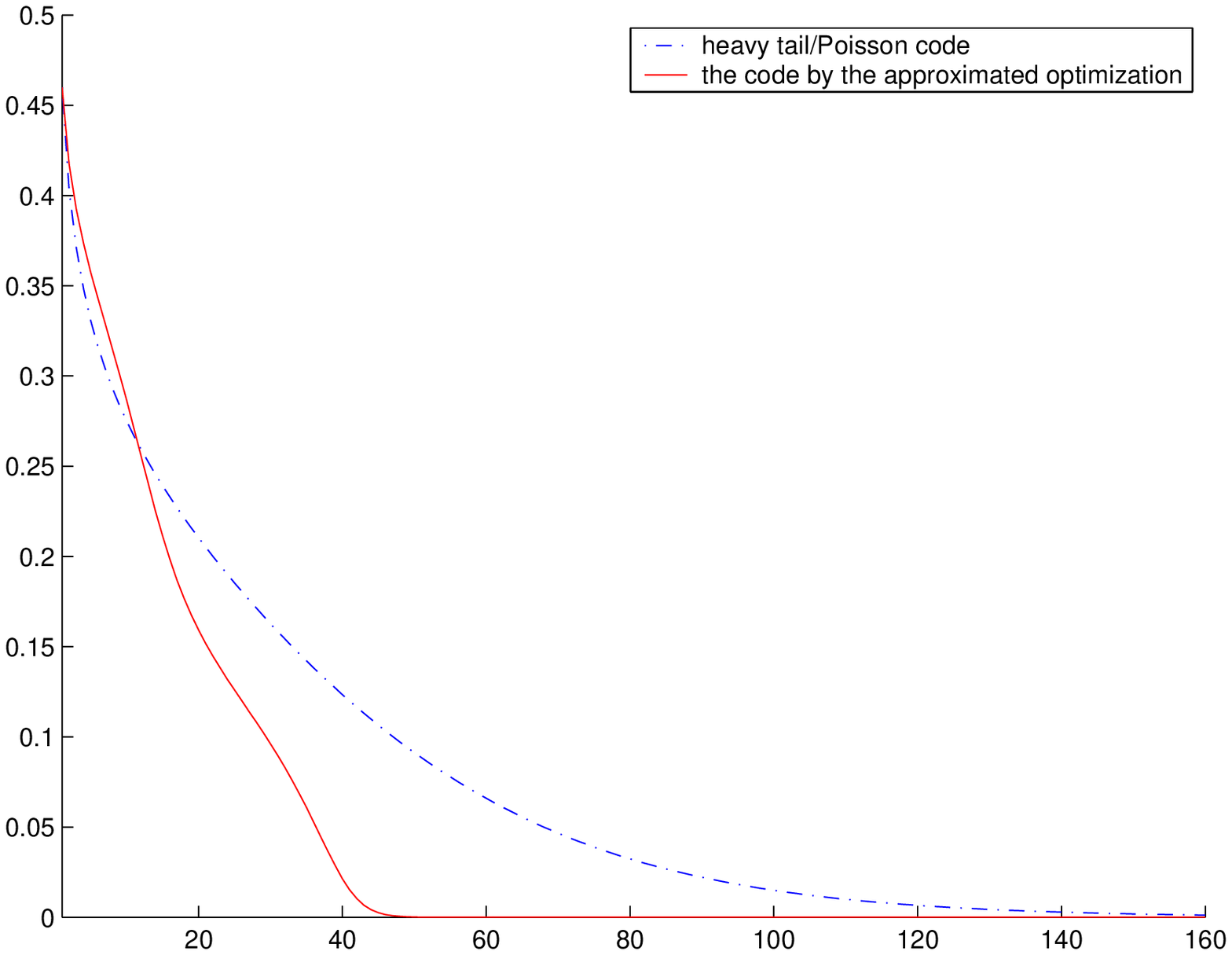}
 \end{center}
 \caption{The message erasure probabilities}
 \label{example4}
\end{figure}

\begin{example}[Example 4:]
In this example, we compare the codes designed by the proposed
approach with the Heavy-tail/Possion codes. We consider a BEC with
channel parameter $\xi=0.46$. Two codes with half rate and maximal
degree $16$ are designed. The left and right degree distributions
of the Heavy-tail/Possion codes are as follows
\cite{Shokrollahi00} :
\begin{align}
\lambda(x)= &
{0.3014x+0.1507x^2+0.1005x^3+0.0753x^4+0.0604x^5+} \nonumber \\
& {0.0502x^6+0.0431x^7+0.0377x^8+0.0335x^9+0.0301x^{10}+}
\nonumber \\
&
{0.0274x^{11}+0.0251x^{12}+0.0232x^{13}+0.0215x^{14}+0.0201x^{15}}
\end{align}
\begin{align}
\rho(x)= & {0.0060x+0.0213x^2+0.0502x^3+0.0887x^4+0.1255x^5+} \nonumber \\
& {0.1479x^6+0.1495x^7+0.1321x^8+0.1039x^9+0.0735x^{10}+}
\nonumber \\
&
{0.0472x^{11}+0.0278x^{12}+0.0151x^{13}+0.0077x^{14}+0.0036x^{15}}
\end{align}
The left and right degree distributions by the proposed design
approach are as follows:
\begin{equation}
\lambda(x)=0.1819x+0.4101x^2+0.0152x^7+0.3928x^{15}
\end{equation}
\begin{equation}
\rho(x)=0.0891x^6+0.9109x^7
\end{equation}
The decoding convergence time $T_{\eta}$, $\eta=10^{-3}$, is $47$
for the code by the proposed approach and $263$ for the
Heavy-tail/Possion codes. The message erasure probabilities of the
two codes by density evolution are show in Fig. \ref{example4}.
The dash-dot curve shows the message erasure probabilities for the
Heavy-tail/Poisson code. The solid curve shows the message erasure
probabilities for the code by the proposed approach.
\end{example}

These numerical results confirm our theoretical results that the
derivatives of $\xi\lambda(1-\rho(1-x))$ with respect to $x$ are
close to $1$ for density-efficient capacity-approaching codes. The
asymptotic approximation $F(\lambda,\rho,\eta)$ is generally
tight. The proposed approach yields codes with good decoding
speed. The optimal codes are right-concentrated.

\section{Conclusions}
\label{section_conclusion}

In this paper, we present a framework for designing LDPC codes
with fast decoding speed. Both the theoretical discussion and
numerical results show that Density-efficient capacity-approaching
codes satisfy the flatness condition. Asymptotically the decoding
convergence time $T_\eta$ can be approximated by
$F(\lambda,\rho,\eta)$. The asymptotic approximation is generally
tight for practical scenarios. The optimal degree distributions in
the sense of decoding speed are right-concentrated.

\useRomanappendicesfalse
\appendices

\section{The proof of Proposition \ref{pro1}}
\label{proof_pro1}

\begin{proof}
The lower bound of $(-1)[\rho(1-x)]'$ follows from the fact that
\begin{eqnarray}
(-1)[\rho(1-x)]' & = & \sum_{j=2}^{k_ca}\rho_j(j-1)(1-x)^{j-2}
\nonumber \\
& \geq &
\sum_{j=2}^{k_ca}\rho_j(1-x)^{j-2} \nonumber \\
& \geq & \sum_{j=2}^{k_ca}\rho_j(1-x)^{j-1}=\rho(1-x) \nonumber \\
\end{eqnarray}
Note that the maximal right degree is bounded by $k_ca$. The upper
bound follows from
\begin{eqnarray}
(-1)[\rho(1-x)]' & = & \sum_{j=2}^{k_ca}\rho_j(j-1)(1-x)^{j-2}
\nonumber \\
& \leq & (k_ca)\sum_{j=2}^{k_ca}\rho_j(1-x)^{j-2} \nonumber \\
& \leq &
\frac{k_ca}{1-x}\sum_{j=2}^{k_ca}\rho_j(1-x)^{j-1} \nonumber \\
& \leq & \frac{k_ca}{1-\xi}\sum_{j=2}^{k_ca}\rho_j(1-x)^{j-1}
\nonumber \\
& = & \frac{k_ca}{1-\xi}\rho(1-x)
\end{eqnarray}
\end{proof}

\section{The proof of Lemma
\ref{lemma_bound_integral_of_rho}}
\label{proof_lemma_bound_integral_of_rho}
\begin{proof}
Using the change of variable $x=1-v$, we have
\begin{equation}
\int_{0}^{1}\rho(1-x)dx=\int_{0}^{1}\rho(v)dv=b
\end{equation}
Using the change of variable $x=\xi \lambda(u)$, we have
\begin{equation}
\int_{0}^{\xi} \lambda^{-1} (x/\xi)dx =  \int_{0}^{1} \xi
u\lambda'(u) du = \xi -  \frac{b\xi}{\xi+\Delta R}
\end{equation}
The lemma follows.
\end{proof}

\section{ The Proof of Lemma \ref{lemma_rho_lemma_diff_bound}}

\begin{figure}[hbt]
\begin{center}
 \includegraphics[width=5in]{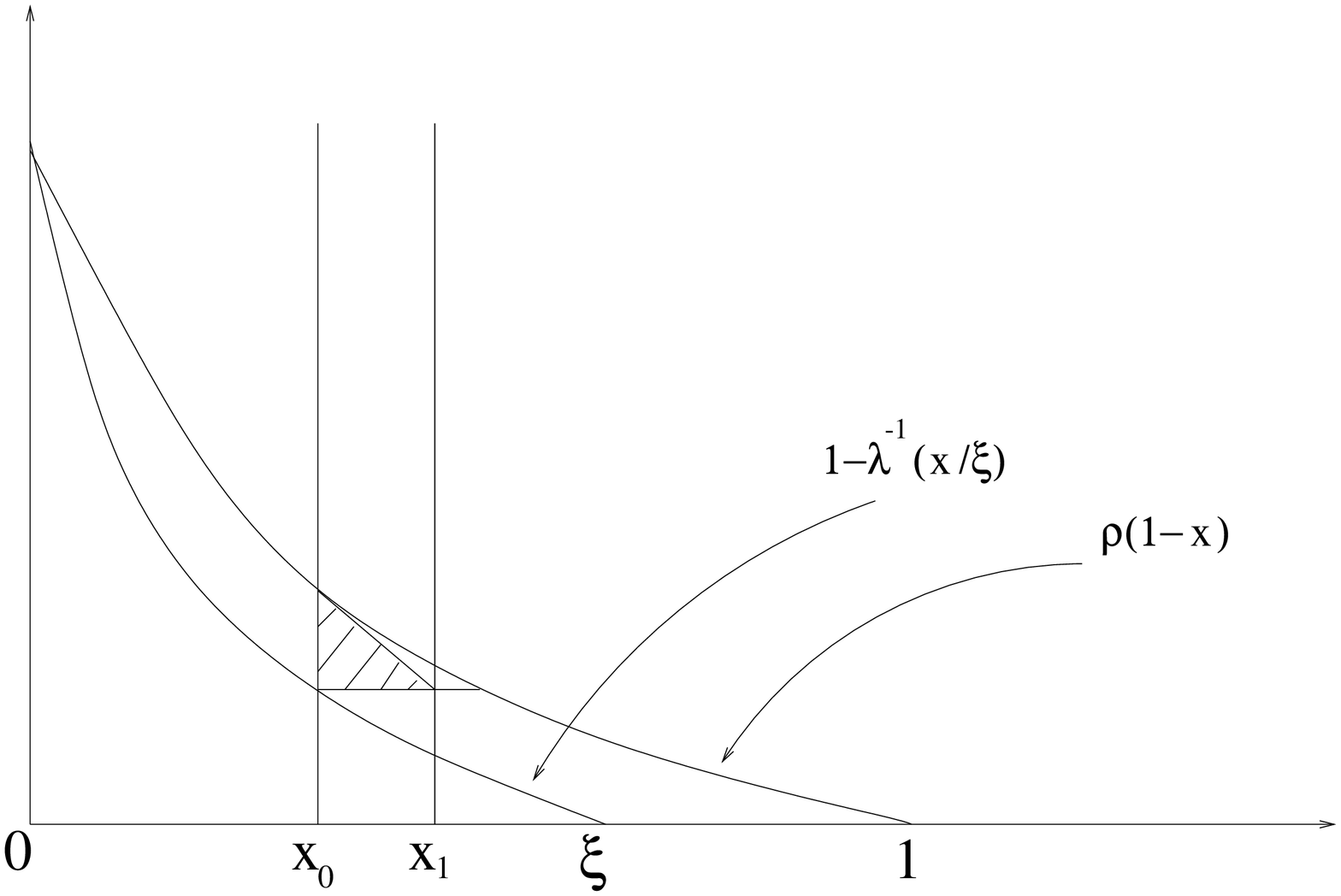}
 \end{center}
 \caption{The geometrical interpretation of $x_0$ and $x_1$}
 \label{proof_demo1}
\end{figure}

\begin{figure}[hbt]
\begin{center}
 \includegraphics[width=5in]{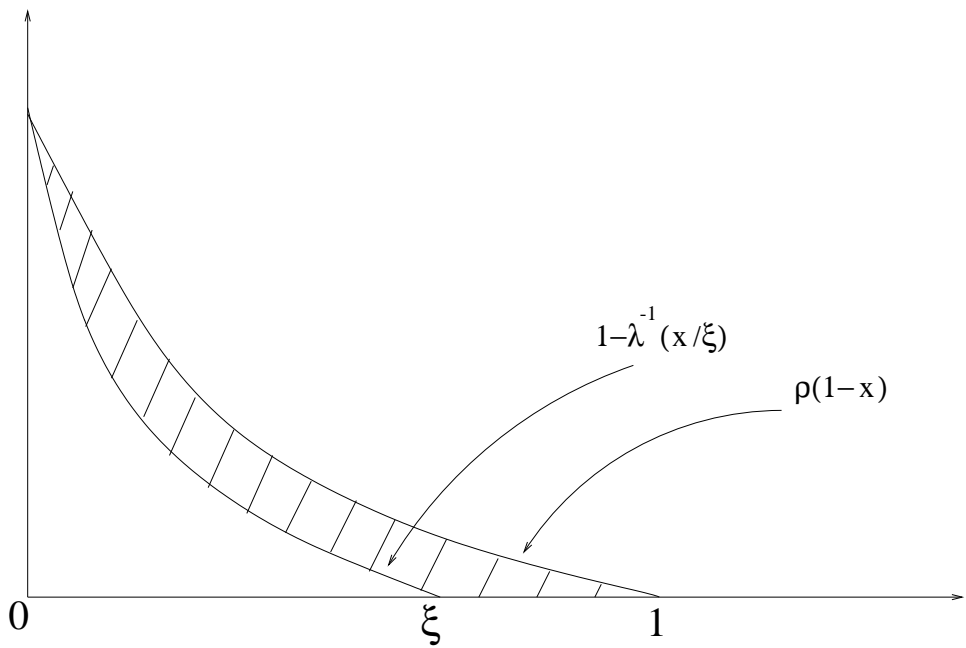}
 \end{center}
 \caption{The geometrical interpretation}
 \label{proof_demo2}
\end{figure}

\begin{proof}
Let us denote $x$ by $x_0$ for convenience and define
\begin{equation}
c\triangleq 1-\frac{1-\lambda^{-1}(x/\xi)} {\rho(1-x)}
\end{equation}
\begin{equation}
x_1\triangleq x_0-\frac{c\rho(1-x_0)}{[\rho(1-x_0)]'}
\end{equation}
The geometric meaning of $x_1$ is shown in Fig. \ref{proof_demo1}.
The number $x_1$ is the $x$ coordinate of the intersection point
of the horizontal line $y=1-\lambda^{-1}(x_0/\xi)$ and the
straight line tangent to the curve $y=\rho(1-x)$ at the point
$(x_0,\rho(1-x_0))$.

The shadowed region in Fig. \ref{proof_demo1} is smaller than the
shadowed region in Fig. \ref{proof_demo2}. The area of the
shadowed region in Fig. \ref{proof_demo1} is
\begin{equation}
\frac{(-1)c^2\left[\rho(1-x_0)\right]^2}{2[\rho(1-x_0)]'}
\end{equation}
The area of the shadowed region in Fig. \ref{proof_demo2} is
\begin{equation}
\int_{0}^{1}\rho(1-x)dx-
\int_{0}^{\xi}\left[1-\lambda^{-1}(x/\xi)\right]dx=\frac{b\Delta
R}{\xi+\Delta R}
\end{equation}
Hence
\begin{equation}
c^2\leq \frac{-2b\Delta R[\rho(1-x_0)]'}{(\xi+\Delta
R)\left[\rho(1-x_0)\right]^2}
\end{equation}
We further bound $(-1)[\rho(1-x_0)]'$ as in Proposition
\ref{pro1}. This gives us the following bound
\begin{equation}
c\leq \sqrt{\frac{2k_c\Delta R}{(\xi+\Delta R)(1-\xi)\rho(1-x_0)}}
\end{equation}
The lemma follows.
\end{proof}

\section{ The Proof of lemma \ref{lemma_bound_recursive_eqn}}
\begin{proof}
We can write $[\lambda(1-\rho(1-x))]''$ as follows:
\begin{align}
 & {[\lambda(1-\rho(1-x))]'' =} \notag \\
& {\left[\sum_{i= 3}^{k_va}\lambda_i
(i-1)(i-2)\left[1-\sum_j\rho_j(1-x)^{j-1}\right]^{i-3}\right]
\times } \notag \\
& {\left[\sum_j\rho_j(j-1)(1-x)^{j-2}\right]^2 - } \notag \\
& { \left[\sum_i\lambda_i(i-1) \times
\left[1-\sum_j\rho_j(1-x)^{j-1}\right]^{i-2}\right] \times }
\notag \\
& { \left[\sum_{j=3}^{k_ca}\rho_j
 (j-1)(j-2)(1-x)^{j-3}\right] } \label{long_eqn_1}
\end{align}
Hence, we have the following bound:
\begin{align}
 & {\left|[\lambda(1-\rho(1-x))]''\right| \leq} \notag \\
& {\left[\sum_{i= 3}^{k_va}\lambda_i
(i-1)(i-2)\left[1-\sum_j\rho_j(1-x)^{j-1}\right]^{i-3}\right]
\times } \notag \\
& {\left[\sum_j\rho_j(j-1)(1-x)^{j-2}\right]^2 + } \notag \\
& { \left[\sum_i\lambda_i(i-1) \times
\left[1-\sum_j\rho_j(1-x)^{j-1}\right]^{i-2}\right] \times }
\notag \\
& { \left[\sum_{j=3}^{k_ca}\rho_j
 (j-1)(j-2)(1-x)^{j-3}\right] } \label{long_eqn_1}
\end{align}

Applying the following bounding,
\begin{equation}
(i-1)<k_va
\end{equation}
\begin{equation}
(i-2)<k_va
\end{equation}
\begin{equation}
(j-1)<k_ca
\end{equation}
\begin{equation}
(j-2)<k_ca
\end{equation}
we have
\begin{align}
&{ \left|[\lambda(1-\rho(1-x))]''\right|
 \leq  k_v^2 k_c^2 a^4 \times } \notag \\
&{ \left[\sum_{i= 3}^{k_va}\lambda_i
\left[1-\sum_j\rho_j(1-x)^{j-1}\right]^{i-3}\right]
\left[\sum_j\rho_j(1-x)^{j-2}\right]^2 } \notag \\
& { +  k_vk_c^2 a^3\left[\sum_i\lambda_i
 \left[1-\sum_j\rho_j(1-x)^{j-1}\right]^{i-2}\right] } \times \notag \\
&{ \left[\sum_{j= 3}^{k_ca}\rho_j
 (1-x)^{j-3}\right] } \label{long_eqn_2}
\end{align}

Note that
\begin{equation}
\left[\sum_{i= 3}^{k_va}\lambda_i
\left[1-\sum_j\rho_j(1-x)^{j-1}\right]^{i-3}\right]\leq 1
\end{equation}
\begin{equation}
\left[\sum_i\lambda_i
 \left[1-\sum_j\rho_j(1-x)^{j-1}\right]^{i-2}\right] \leq 1
\end{equation}
we have
\begin{align}
& { \left|[\lambda(1-\rho(1-x))]''\right|  \leq  k_v^2 k_c^2 a^4
\left[\sum_j\rho_j(1-x)^{j-2}\right]^2+ } \notag \\
&{ k_v k_c^2 a^3\left[\sum_{j= 3}^{k_ca}\rho_{j}
 (1-x)^{j-3}\right] }\label{long_eqn_3}
\end{align}

Further apply the following upper bounds for
$\sum_j\rho_j(1-x)^{j-2}$ and $\sum_{j=3}^{k_ca}\rho_{j}
(1-x)^{j-3}$
\begin{equation}
\sum_j\rho_j(1-x)^{j-2}= \frac{\sum_j\rho_j(1-x)^{j-1}}{1-x}\leq
\frac{\rho(1-x)}{1-\xi}
\end{equation}
\begin{equation}
\sum_{j= 3}^{k_ca}\rho_{j}
 (1-x)^{j-3} \leq \frac{\sum_{j= 2}^{k_ca}\rho_{j}
 (1-x)^{j-1}}{(1-x)^2}\leq \frac{\rho(1-x)}{(1-\xi)^2}
\end{equation}
we have
\begin{equation}
\left|[\lambda(1-\rho(1-x))]''\right| \leq
\frac{k_v^2k_c^2\rho(1-x)^2a^4}{(1-\xi)^2}+
\frac{k_vk_c^2\rho(1-x)a^3}{(1-\xi)^2}
\end{equation}
\end{proof}

\section{The Proof of lemma \ref{lemma_bound_ratio_derivatives}}

\begin{proof}

Let $x_0\in(b^5,\xi-b^2)$. Let $x\in(0,\xi)$, $x\neq x_0$, we have
the following Taylor series expansion
\begin{align}
& { 1-\lambda^{-1}(x/\xi)-\rho(1-x)  =
1-\lambda^{-1}(x_0/\xi)-\rho(1-x_0)+ } \notag \\
& { \left\{\left[1-\lambda^{-1}(x_0/\xi)\right]'
 -  \left[\rho(1-x_0)\right]'\right\}(x-x_0)+ } \notag \\
 & { \left\{\left[1-\lambda^{-1}(\zeta/\xi)\right]''
-\left[\rho(1-\zeta)\right]''\right\} \frac{(x-x_0)^2}{2} }
\label{long_eqn_4}
\end{align}
where $\zeta$ is a real number between $x_0$ and $x$. According to
the hypotheses, $\xi\lambda(1-\rho(1-x))<x$. This implies
$1-\lambda^{-1}(x/\xi)-\rho(1-x)<0$, and
\begin{align}
& {1-\lambda^{-1}(x_0/\xi)-\rho(1-x_0)+ } \notag \\
& { \left\{\left[1-\lambda^{-1}(x_0/\xi)\right]'
 -  \left[\rho(1-x_0)\right]'\right\}(x-x_0)+ } \notag \\
 & { \left\{\left[1-\lambda^{-1}(\zeta/\xi)\right]''
-\left[\rho(1-\zeta)\right]''\right\} \frac{(x-x_0)^2}{2}<0 }
\label{long_eqn_4}
\end{align}

Note that $[1-\lambda^{-1}(\zeta/\xi)]''>0$, we therefore have the
following more convenient inequality:
\begin{align}
& { 1-\lambda^{-1}(x_0/\xi)-\rho(1-x_0)+ } \notag \\
& { \left\{\left[1-\lambda^{-1}(x_0/\xi)\right]'
-\left[\rho(1-x_0)\right]'\right\}(x-x_0) } \notag \\
& { -\left[\rho(1-\zeta)\right]'' \frac{(x-x_0)^2}{2}<0 }
\label{long_eqn_5}
\end{align}

Set $x=x_0+b^2$ in the above inequality, with a little algebra we
have
\begin{equation}
\label{lemma_37_eqn_lower_bound} \left\{
\frac{[1-\lambda^{-1}(x_0/\xi)]'}{[\rho(1-x_0)]'}-1 \right\} \geq
\frac{b^2[\rho(1-\zeta)]''}{2[\rho(1-x_0]'} +
\frac{\left[\lambda^{-1}(x_0/\xi)+\rho(1-x_0)-1\right]}{b^2[\rho(1-x_0)]'}
\end{equation}
To prove the lower bound in the lemma, we will bound the two terms
in the right hand side of Eqn. \ref{lemma_37_eqn_lower_bound}
separately.

We bound the first term as follows. Since $[\rho(1-x)]''$ is a
monotonous decreasing function,
\begin{equation}
[\rho(1-\zeta)]''  \leq  [\rho(1-x_0)]''
\end{equation}
Apply the bound in Proposition \ref{pro2}, we hve
\begin{eqnarray}
[\rho(1-\zeta)]''  \leq \frac{(-1)k_ca}{(1-\xi)}[\rho(1-x_0)]'
\end{eqnarray}
Thus, the first term in the right hand side of Eqn.
\ref{lemma_37_eqn_lower_bound} can be lower bounded
\begin{equation}
\frac{b^2[\rho(1-\zeta)]''}{2[\rho(1-x_0]'}  \geq
\frac{-k_cb}{2(1-\xi)} \label{lemma_37_result2}
\end{equation}

We bound the second term in the right hand side of Eqn.
\ref{lemma_37_eqn_lower_bound} as follows. The second term in the
right hand side of Eqn. \ref{lemma_37_eqn_lower_bound} can be
rewritten as follows:
\begin{align}
& {\frac{\rho(1-x_0)-1+\lambda^{-1}(x_0/\xi)}{b^2[\rho(1-x_0)]'}
 =} \notag \\
 & { (-a^2)\left\{\frac{\rho(1-x_0)-1+\lambda^{-1}(x_0/\xi)}{\rho(1-x_0)}\right\}
\left\{\frac{\rho(1-x_0)}{(-1)[\rho(1-x_0)]'}\right\} }
\end{align}
According to Lemma \ref{lemma_rho_lemma_diff_bound},
\begin{equation}
\frac{\rho(1-x_0)-1+\lambda^{-1}(x_0/\xi)}{\rho(1-x_0)} \leq
\sqrt{k_c}B(\Delta R, b,x_0)
\end{equation}
According to Proposition \ref{pro1}
\begin{equation}
\left\{\frac{\rho(1-x_0)}{-[\rho(1-x_0)]'}\right\}\leq 1
\end{equation}
Hence, the second term in the right hand side of Eqn.
\ref{lemma_37_eqn_lower_bound} can be lower bounded as follows.
\begin{equation}
\frac{\rho(1-x_0)-1+\lambda^{-1}(x_0/\xi)}{[\rho(1-x_0)]'b^2} \geq
-\sqrt{k_c}B(\Delta R,b,x)a^2 \label{lemma_37_result1}
\end{equation}

Substituting Eqns. \ref{lemma_37_result1} and
\ref{lemma_37_result2} into Eqn. \ref{lemma_37_eqn_lower_bound}
gives the lower bound in the lemma.

We will prove the upper bound in the lemma. Set $x=x_0-b^5$ in
Eqn. \ref{long_eqn_5}, with a little algebra we have
\begin{align}
\frac{[1-\lambda^{-1}(x_0/\xi)]'}{[\rho(1-x_0)]'}-1   \leq &{
\frac{1-\lambda^{-1}(x_0/\xi)-\rho(1-x_0)}{b^5[\rho(1-x_0)]'} }
\notag \\
& {+  \frac{-b^5[\rho(1-\zeta)]''}{2[\rho(1-x_0)]'}}
\label{eqn_bound1}
\end{align}
Since of Lemma \ref{lemma_rho_lemma_diff_bound} and Proposition
\ref{pro1}, the first term at the right hand side of Eqn.
\ref{eqn_bound1} can be bounded
\begin{align}
 \frac{1-\lambda^{-1}(x_0/\xi)-\rho(1-x_0)}{[\rho(1-x_0)]'b^5}
  \leq &  {\frac{a^5\sqrt{k_c}B(\Delta R,b,x)\rho(1-x_0)}{(-1)[\rho(1-x_0)]'}}
 \notag \\
  \leq & { \sqrt{k_c}B(\Delta R,b,x)a^5}
  \label{lemma_37_result3}
\end{align}
Note that the maximal right degree is bounded by $k_ca$. Hence
\begin{align}
[\rho(1-\zeta)]''  \leq & {
[\rho(1-x_0)]''\left[\frac{1-\zeta}{1-x_0}\right]^{k_ca-2} }
\notag \\
 \leq & {
[\rho(1-x_0)]''\left[\frac{1-x_0+b^5}{1-x_0}\right]^{k_ca-2} }
\notag \\
 \leq  & {[\rho(1-x_0)]''\left[1+\frac{b^5}{1-\xi}\right]^{k_ca}}
\end{align}
By bounding $[\rho(1-\zeta)]''$ as in Proposition \ref{pro2}, the
second term at the right hand side of Eqn. \ref{eqn_bound1} can be
bounded by
\begin{equation}
(-1)\frac{[\rho(1-\zeta)]''b^5}{2[\rho(1-x_0)]'} \leq
\frac{k_cb^4}{2(1-\xi)}  \left(1+\frac{b^5}{1-\xi}\right)^{k_ca}
\label{lemma_37_result4}
\end{equation}
Substituting Eqns. \ref{lemma_37_result3} and
\ref{lemma_37_result4} into Eqn. \ref{eqn_bound1} gives the upper
bound in the lemma.
\end{proof}

\section{ The Proof of lemma
\ref{lemma_bound_derivatives}}

\begin{proof}
Notice that
\begin{equation}
[\xi\lambda(1-\rho(1-x_0))]' =\frac{[\rho(1-x_0)]'}
{[1-\lambda^{-1} (x_1/\xi)]'}
\end{equation}
Since $[\rho(1-x)]'$ and $[1-\lambda^{-1}(x/\xi)]'$ are
monotonously increasing,
\begin{equation}
\frac{[\rho(1-x_1)]'}{[1-\lambda^{-1}(x_1/\xi)]'} \geq
\frac{[\rho(1-x_0)]'}{[1-\lambda^{-1}(x_1/\xi)]'}
\end{equation}
\begin{equation}
\frac{[\rho(1-x_0)]'}{[1-\lambda^{-1}(x_0/\xi)]'} \geq
\frac{[\rho(1-x_0)]'}{[1-\lambda^{-1}(x_1/\xi)]'}
\end{equation}
The lemma follows.
\end{proof}

\section{The Proof of lemma
\ref{lemma_bound_two_curve_y}}

\begin{figure}[hbt]
\begin{center}
 \includegraphics[width=5in]{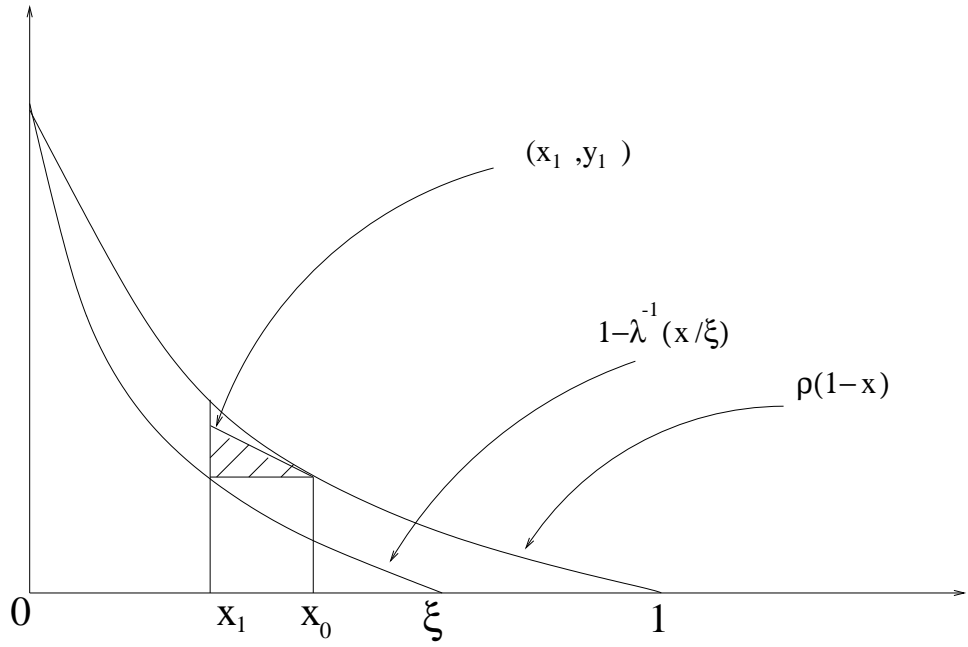}
 \end{center}
 \caption{The geometrical interpretation of $x_0$ and $x_1$.}
 \label{proof_demo4}
\end{figure}

\begin{proof}

Denote $x$ by $x_0$ for convenience. Define
\begin{equation}
x_1=\xi\lambda(1-\rho(1-x_0))
\end{equation}
\begin{equation}
y_0=\rho(1-x_0)
\end{equation}
\begin{equation}
y_1=y_0-(x_0-x_1)[\rho(1-x_0)]'
\end{equation}
The geometric meaning of $x_0$, $x_1$, $y_0$, and $y_1$ is shown
in Fig. \ref{proof_demo4}. The point $(x_1,y_1)$ is the
intersection of the vertical straight line $x=x_1$ and the
straight line tangent to the curve $\rho(1-x)$ at the point
$(x_0,y_0)$.

The area of the shadowed region in Fig. \ref{proof_demo4} is
\begin{equation}
\frac{-1}{2}[\rho(1-x_0)]'(x_0-x_1)^2
\end{equation}
The shadowed region in Fig. \ref{proof_demo4} is smaller than the
shadowed region in Fig. \ref{proof_demo2},
\begin{equation}
\frac{-1}{2}[\rho(1-x_0)]'(x_0-x_1)^2 \leq \frac{b \Delta
R}{\xi+\Delta R}
\end{equation}
The lemma follows.
\end{proof}

\section{The Proof of Lemma
\ref{lemma_boundary}} \label{proof_lemma_boundary}

\begin{figure}[hbt]
\begin{center}
 \includegraphics[width=5in]{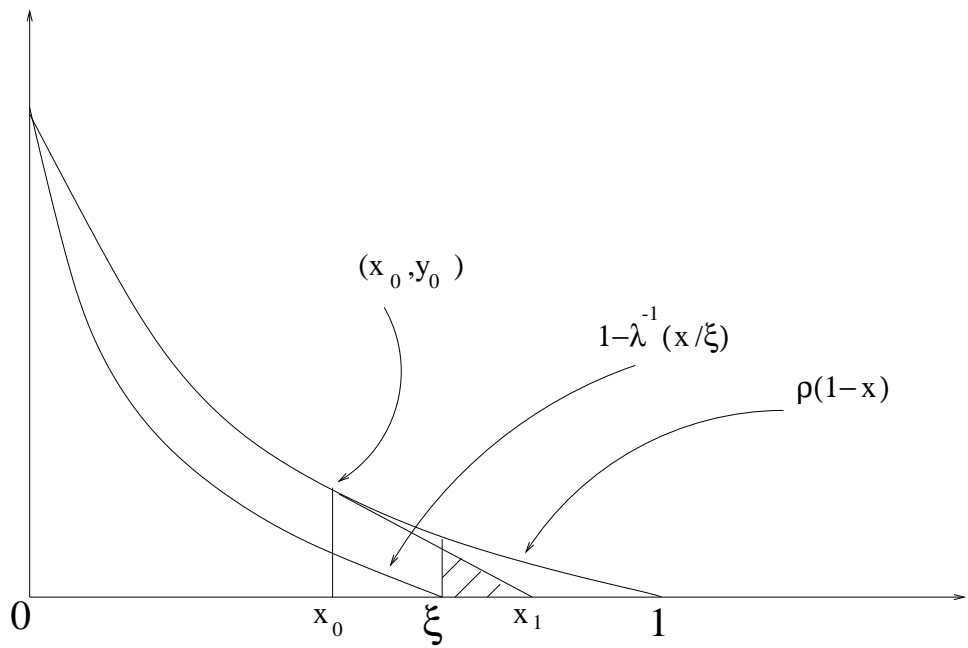}
 \end{center}
 \caption{The geometrical interpretation of $x_1$}
 \label{proof_demo3}
\end{figure}

\begin{proof}
Define
\begin{equation}
x_0=\xi-b^2
\end{equation}
\begin{equation}
y_0=\rho(1-x_0)
\end{equation}
\begin{equation}
x_1=x_0+\frac{y_0}{(-1)[\rho(1-x_0)]'}
\end{equation}
Since $b<(1-\xi)/k_c$,
\begin{align}
\label{bound_for_x1} x_1   = &
{x_0+\frac{y_0}{(-1)[\rho(1-x_0)]'}\geq x_0+
\frac{(1-\xi)y_0 b}{k_c\rho(1-x_0)}} \notag \\
 \geq & {x_0+\frac{(1-\xi)b}{k_c} >  \xi}
\end{align}
The geometric meaning of $x_1$ is shown in Fig. \ref{proof_demo3}.
The point $(x_1,0)$ is the intersection of the $x$-axis and the
straight line tangent to the curve $y=\rho(1-x)$ at the point
$(x_0,y_0)$.

The shadowed region in Fig. \ref{proof_demo3} is smaller than the
shadowed region in Fig. \ref{proof_demo2}. For the area of the
shadowed triangle region in Fig. \ref{proof_demo3}, the width is
$x_1-x_0$, the height is $(x_1-\xi)y_0/{x_1-x_0}$, and the area is
\begin{equation}
\frac{y_0(x_1-\xi)^2}{2(x_1-x_0)}
\end{equation}
Since this area is monotonously increase with respect to $x_1$,
applying the bound for $x_1$ in Eqn. \ref{bound_for_x1}, we have
the following lower bound of the area
\begin{equation}
\frac{y_0 b}{2k_c(1-\xi)}\left(1-\xi-k_cb\right)^2
\end{equation}
This lower bound is less than the area of the shadowed region in
Fig. \ref{proof_demo2}
\begin{equation}
\frac{y_0 b}{2k_c(1-\xi)}\left(1-\xi-k_cb\right)^2  \leq
\frac{b\Delta R}{\xi+\Delta R}
\end{equation}
The lemma follows.
\end{proof}

\section{The Proof of Theorem \ref{theorem_flatness_main}}
\label{proof_theorem_flatness_main}

\begin{proof}
The proof is divided into three steps.

\newcounter{step_theo_flat_main}

Step \Roman{proof_steps}: We will define a partition of the
interval $(0,\xi]$.

For any $n$, we partition the interval $(0,\xi]$ into three
subintervals $(0,\zeta_0]$, $(\zeta_0,\zeta_1]$, and
$(\zeta_1,\xi]$, where
\begin{equation}
\zeta_0=2b_n^5
\end{equation}
\begin{equation}
\zeta_1=1-\rho_n^{-1}\left((b_n)^5\right)
\end{equation}

we claim that the partition is well-defined for sufficiently large
$n$. That is, $\zeta_0<\zeta_1$ for sufficiently large $n$. Note
that
\begin{equation}
\rho_n(1-\zeta_0)  =  \rho_n\left(1-2(b_n)^5\right)=
\sum_{j=2}^{k_ca_n}\rho_{nj} \left(1-2(b_n)^5\right)^{j-1}
\end{equation}
Lower bounding $\left[1-2(b_n)^5\right]^{j-1}$ by $
\left[1-2(b_n)^5\right]^{k_ca_n-1}$, we have
\begin{equation}
\rho_n(1-\zeta_0) \geq \left[1-2(b_n)^5\right]^{k_ca_n-1}
\end{equation}
This lower bound of $\rho_n(1-\zeta_0)$ goes to $1$, as $n$ goes
to infinity. Hence
\begin{equation}
\rho_n(1-\zeta_0)\rightarrow 1
\end{equation}
On the other hand,
\begin{equation}
\rho_n(1-\zeta_1)=(b_n)^5\rightarrow 0
\end{equation}
Since $\rho_n(1-x)$ is a monotonously decreasing function, we
conclude that $\zeta_0<\zeta_1$ for sufficiently large $n$.

We claim that $\zeta_1<\xi-(b_n)^2$ for sufficiently large $n$.
According to Lemma \ref{lemma_boundary},
\begin{equation}
\rho_n\left(1-\xi+(b_n)^2\right)=o\left((b_n)^{15}\right)
\end{equation}
while by definition
\begin{equation}
\rho_n(1-\zeta_1)= (b_n)^5
\end{equation}
Hence $\zeta_1<\xi-(b_n)^2$. The claim is proven.

\addtocounter{proof_steps}{1} Step \Roman{proof_steps}: In this
step, we show that the derivative of the function
$\xi\lambda_n(1-\rho_n(1-x))$ converges to $1$ uniformly in the
subinterval $(\zeta_0,\zeta_1)$.

We will show that the function $[\xi\lambda_n(1-\rho_n(1-x))]'$ is
upper bounded and this upper bound goes to $1$ uniformly as $n$
goes to infinity. According to Lemma
\ref{lemma_bound_derivatives},
\begin{equation}
[\xi \lambda_n(1-\rho_n(1-x))]'\leq \frac{[\rho_n(1-x)]'}
{[1-\lambda_n^{-1}(x/\xi)]'}
\end{equation}
According to lemma \ref{lemma_bound_ratio_derivatives},
\begin{equation}
\frac{[1-\lambda_n^{-1}(x/\xi)]'}{[\rho_n(1-x)]'}\geq 1-
\frac{k_cb_n}{2(1-\xi)}-\sqrt{k_c}B(\Delta
R_n,b_n,x)(1-\xi)(a_n)^2
\end{equation}
Also note that
\begin{equation}
\Delta R_n=o\left((b_n)^{15}\right)
\end{equation}
\begin{equation}
B(\Delta R_n,b_n,x)=o\left((b_n)^5\right)
\end{equation}
We conclude that
\begin{eqnarray}
\frac{[1-\lambda_n^{-1}(x/\xi)]'}{[\rho_n(1-x)]'}  \geq  1+O(b_n)
\end{eqnarray}
Therefore, the function $[\xi\lambda_n(1-\rho_n(1-x))]'$ is upper
bounded,
\begin{equation}
[\xi\lambda_n(1-\rho_n(1-x))]'\leq \frac{1}{1+O(b_n)}
\end{equation}
This upper bound goes to $1$ as $n$ goes to infinity.

We claim that for $x\in(\zeta_0,\zeta_1]$,
\begin{equation}
x-\xi\lambda_n(1-\rho_n(1-x))=o\left((b_n)^{5.5}\right)
\end{equation}
Hence for $x\in(\zeta_0,\zeta_1]$,
\begin{equation}
\xi\lambda_n(1-\rho_n(1-x))>(b_n)^5
\end{equation}
for sufficiently large $n$. Denote $\xi\lambda_n(1-\rho_n(1-x))$
by $y$. According to Lemma \ref{lemma_bound_two_curve_y},
\begin{eqnarray}
\Delta x & = & x-y \leq \sqrt{\frac{-2b_n\Delta R_n}{(\xi+\Delta
R_n)[\rho(1-x)]'}}
\end{eqnarray}
Bounding $[\rho_n(1-x)]'$ as in Proposition \ref{pro1}, we have
\begin{eqnarray}
\Delta x & = & x-y  \leq  \sqrt{\frac{2b_n\Delta R_n}{(\xi+\Delta
R_n)[\rho(1-x)]}}
\end{eqnarray}
Note that
\begin{equation}
\rho_n(1-x)\geq \rho_n(1-\zeta_1)=(b_n)^5
\end{equation}
\begin{equation}
\Delta R_n=o\left((b_n)^{15}\right)
\end{equation}
We have
\begin{eqnarray}
\Delta x =o\left((b_n)^{5.5}\right)
\end{eqnarray}
Therefor for sufficiently large $n$,
\begin{equation}
\xi\lambda_n(1-\rho(1-x))=x-\Delta x \geq \zeta_0-\Delta x
>(b_n)^5
\end{equation}

We will show that the function $[\xi\lambda_n(1-\rho_n(1-x))]'$ is
also lower bounded and this lower bound converges to $1$ as $n$
goes to infinity. Note that
\begin{align}
[\xi\lambda_n(1-\rho_n(1-x))]'  = &
{\frac{[\rho_n(1-x)]'}{[1-\lambda_n^{-1}(y/\xi)]'}} \notag \\
 = &
 {\left\{\frac{[\rho_n(1-y)]'}{[1-\lambda_n^{-1}(y/\xi)]'}\right\}
\left\{\frac{[\rho_n(1-x)]'}{[\rho_n(1-y)]'}\right\}}
\end{align}
where $y=\xi\lambda_n(1-\rho_n(1-x))$. For sufficiently large $n$,
we bound the second term as follows:
\begin{equation}
\frac{[\rho_n(1-x)]'}{[\rho_n(1-y)]'}\geq
\left(\frac{1-x}{1-y}\right)^{k_ca_n-1} \geq
\left(\frac{1-x}{1-x+b_n^{5.3}}\right)^{k_ca_n-1}
\end{equation}
We have the following lower bound for
$[\xi\lambda_n(1-\rho_n(1-x))]'$
\begin{equation}
[\xi\lambda_n(1-\rho_n(1-x))] \geq
\frac{[\rho_n(1-y)]'}{[1-\lambda_n^{-1}(y/\xi)]'}
\left(\frac{1-x}{1-x+b_n^{5.3}}\right)^{k_ca_n-1}
\end{equation}
Since $y\geq (b_n)^5$, according to Lemma
\ref{lemma_bound_ratio_derivatives} we have,
\begin{equation}
\frac{[\rho_n(1-y)]'} {[1-\lambda_n^{-1}(y/\xi)]'}\rightarrow 1
\mbox{ as }n\rightarrow \infty
\end{equation}
Also
\begin{equation}
\left(\frac{1-x}{1-x+b_n^{5.3}}\right)^{k_ca_n-1}\rightarrow 1
\end{equation}
as $n\rightarrow \infty$. We conclude that this lower bound for
$[\xi\lambda_n(1-\rho_n(1-x))]'$ converges to $1$ as $n$ goes to
infinity.

From the above, we conclude that $[\xi\lambda_n(1-\rho_n(x))]'$
converges to $1$ uniformly for $x\in(\zeta_0,\zeta_1]$ as $n$ goes
to infinity.

\addtocounter{proof_steps}{1} Step \Roman{proof_steps}: In this
step, we show that the function $[\xi\lambda_n(1-\rho_n(1-x))]'$
also converges to $1$ uniformly in the subintervals $(0,\zeta_0)$
and $(\zeta_1,\xi]$.

For $x\in(0,\zeta_0)$, according to Lemma
\ref{lemma_bound_recursive_eqn},
\begin{equation}
\left|[\xi\lambda_n(1-\rho_n(1-x))]''\right|\leq \frac{k_v^2 k_c^2
\left[\rho_n(1-x)\right]^2a_n^4}{(1-\xi)^2}  + \frac{k_vk_c^2
\rho_n(1-x)a_n^3}{1-\xi}  \leq O(a_n^4)
\end{equation}
while the length of this interval is $2b_n^5$. Hence
$[\xi\lambda_n(1-\rho_n(1-x))]'$ converges to $1$ uniformly for
$x\in(0,\zeta_0)$.

For $x\in(\zeta_1,\xi)$, according to Lemma
\ref{lemma_bound_recursive_eqn},
\begin{equation}
\left|[\xi\lambda_n(1-\rho_n(1-x))]''\right|\leq \frac{k_v^2 k_c^2
\left[\rho_n(1-x)\right]^2a_n^4}{(1-\xi)^2}  + \frac{k_vk_c^2
\rho_n(1-x)a_n^3}{1-\xi}   = O(b_n^2);
\end{equation}
while the length of this interval is bounded by $1$. Hence
$[\xi\lambda_n(1-\rho_n(1-x))]'$ also converges to $1$ uniformly
in the interval $(\zeta_2,\xi)$. The theorem is proven.
\end{proof}

\section{The Proof of Theorem
 \ref{theorem_asym_approx}}
\label{proof_theorem_asym_approx}
\begin{proof}
The proof is divided into four steps.

\setcounter{proof_steps}{1} Step \Roman{proof_steps}: in this
step, we define a partition of the interval $(0,\xi)$.

According to Theorem \ref{theorem_flatness_main}, the derivative
of $\xi \lambda_n(1-\rho_n(1-x))$ converge uniformly to $1$ for
$x\in(0,\xi]$ as $n$ goes to infinity. There exists an
$\epsilon_n$ such that
\begin{equation}
\left|\left[x-\xi\lambda_n(1-\rho_n(1-x))\right]'\right|\leq
\epsilon_n
\end{equation}
for $x\in(0,\xi)$ and $\epsilon_n\rightarrow 0$ as $n\rightarrow
\infty$.

We partition the interval $(\eta,\xi)$ into subintervals
$(x_0,x_1)$, $(x_1,x_2)$, $\cdots$, $(x_k,x_{k+1})$, $\cdots$,
$(x_{m-1},x_m)$. The real numbers $x_i$ are recursively defined as
follows:
\begin{itemize}
\item Step (a), set $k=0$, $x_0=\eta$.

\item Step (b), set
\begin{eqnarray}
\Delta x_k=\min \left \{
\frac{1}{\sqrt{\epsilon_n}}[x_k-\xi\lambda(1-\rho(1-x_k))],
\frac{\xi}{2} \right \} \nonumber \\
\end{eqnarray}
\begin{equation}
x_{k+1}=x_k+\Delta x_k
\end{equation}

\item Step (c), if $x_{k+1}\geq \xi$, set $x_{k+1}=\xi$, $m=k+1$
and stop. Otherwise, set $k=k+1$, go to step (b).
\end{itemize}

\addtocounter{proof_steps}{1} Step \Roman{proof_steps}: in this
step, we show an upper bound for
\begin{equation}
\frac{\max_{x\in(x_k,x_{k+1})}[x-\xi\lambda_n(1-\rho_n(1-x))]}
{\min_{x\in(x_k,x_{k+1})}[x-\xi\lambda_n(1-\rho_n(1-x))]}
\end{equation}

 For each interval $(x_k,x_{k+1})$, the length of the
interval is at most
\begin{equation}
\frac{1}{\sqrt{\epsilon_n}}[x_k-\xi\lambda_n(1-\rho_n(1-x_k))]
\end{equation}
Hence,
\begin{eqnarray}
\min_{x\in(x_k,x_{k+1})}[x-\xi\lambda_n(1-\rho_n(1-x))] \\
 \geq  [x_k-\xi\lambda_n(1-\rho_n(1-x_k))] [1-\sqrt{\epsilon_n}]
\end{eqnarray}
\begin{eqnarray}
\max_{x\in(x_k,x_{k+1})}[x-\xi\lambda_n(1-\rho_n(1-x))] \nonumber
\\ \leq [x_k-\xi\lambda_n(1-\rho_n(1-x_k))] [1+\sqrt{\epsilon_n}]
\end{eqnarray}
Therefore, we have the following upper bound
\begin{equation}
\frac{\max_{x\in(x_k,x_{k+1})}[x-\xi\lambda_n(1-\rho_n(1-x))]}
{\min_{x\in(x_k,x_{k+1})}[x-\xi\lambda_n(1-\rho_n(1-x))]} \leq
\frac{1+\sqrt{\epsilon_n}}{1-\sqrt{\epsilon_n}}
\label{max_min_ratio_bound}
\end{equation}

\addtocounter{proof_steps}{1} Step \Roman{proof_steps}: in this
step, we show lower and upper bounds for
$F(\lambda,\rho,\eta)/T_\eta$.

Denote the message erasure probability at $l$-th iteration by
$P_e^{(l)}$. Let $T_k$ be the number of $P_e^{(l)}$ such that
$P_{e}^{(l)}\in [x_{k},x_{k+1})$. Note that the message erasure
probability decreases at least
\begin{equation}
\min_{x\in [x_{k},x_{k+1})} [x-\xi\lambda_n(1-\rho_n(1-x))]
\end{equation}
and at most
\begin{equation}
\max_{x\in [x_{k},x_{k+1})} [x-\xi\lambda_n(1-\rho_n(1-x))]
\end{equation} during each iteration. Hence, the following
inequalities hold
\begin{equation}
\Delta x_k \geq (T_k-1)\min_{x\in [x_{k},x_{k+1})}
[x-\xi\lambda_n(1-\rho_n(1-x))]
\end{equation}
\begin{equation}
\Delta x_k\leq (T_k+1)\max_{x\in [x_{k},x_{k+1})}
[x-\xi\lambda_n(1-\rho_n(1-x))]
\end{equation}
It follows that
\begin{eqnarray}
T_k & \geq & \frac{\Delta x}
{\max_{x\in(x_{k},x_{k+1})}[x-\xi\lambda_n(1-\rho_n(1-x))]}-1
\nonumber \\
\end{eqnarray}
\begin{eqnarray}
T_k & \leq & \frac{\Delta x}{ \min_{x\in (x_{k},x_{k+1})}
[x-\xi\lambda_n(1-\rho_n(1-x))]}+1 \nonumber \\
\end{eqnarray}
According to the bounds in Eqn. \ref{max_min_ratio_bound}, we have
\begin{eqnarray}
T_k & \geq & \frac{1-\sqrt{\epsilon_n}}{1+\sqrt{\epsilon_n}}
\max_{x\in(x_{k},x_{k+1})} \frac{\Delta x}
{[x-\xi\lambda_n(1-\rho_n(1-x))]}-1
\nonumber \\
\label{eqn_t_k_bound1}
\end{eqnarray}
\begin{eqnarray}
T_k & \leq & \frac{1+\sqrt{\epsilon_n}}{1-\sqrt{\epsilon_n}}
\min_{x\in (x_{k},x_{k+1})} \frac{\Delta x}{
[x-\xi\lambda_n(1-\rho_n(1-x))]}+1 \nonumber \\
\end{eqnarray}
Note that $T_{\eta}=\sum_{k=1}^{m}T_k$, we have
\begin{equation}
T_\eta\geq \frac{1-\sqrt{\epsilon_n}}{1+\sqrt{\epsilon_n}}
\int_{\eta}^{\xi} \frac{1}{x-\xi\lambda_n(1-\rho_n(1-x))} dx-m
\end{equation}
\begin{equation}
T_\eta\leq \frac{1+\sqrt{\epsilon_n}}{1-\sqrt{\epsilon_n}}
\int_{\eta}^{\xi} \frac{1}{x-\xi\lambda_n(1-\rho_n(1-x))} dx+m
\end{equation}
Therefore, we have the following bounds for the ratio
$F(\lambda_n,\rho_n,\eta)/T_\eta$
\begin{equation}
\left(1+\frac{m}{T_\eta}\right)
\frac{1+\sqrt{\epsilon_n}}{1-\sqrt{\epsilon_n}} \geq
\frac{F(\lambda_n,\rho_n,\eta)}{T_\eta} \geq
\left(1-\frac{m}{T_\eta}\right)
\frac{1-\sqrt{\epsilon_n}}{1+\sqrt{\epsilon_n}} \label{eqn_temp2}
\end{equation}

\addtocounter{proof_steps}{1} Step \Roman{proof_steps}: in this
step, we show that the lower and upper bound in the last step all
converges to $1$ as $n$ goes to infinity.

Note that
\begin{equation}
\lim_{n\rightarrow
\infty}\frac{1+\sqrt{\epsilon_n}}{1-\sqrt{\epsilon_n}}=
\lim_{n\rightarrow
\infty}\frac{1-\sqrt{\epsilon_n}}{1+\sqrt{\epsilon_n}}=1
\end{equation}
It suffices to show that $m/T_\eta\rightarrow 0$ as $n\rightarrow
\infty$.

We claim that
\begin{equation}
\min_{1\leq k\leq m-1}T_k\rightarrow \infty,\,\,\,
\mbox{as}\,\,\,n \rightarrow \infty
\end{equation}
According to Eqn. \ref{eqn_t_k_bound1}
\begin{eqnarray}
T_k & \geq &
\left(\frac{1-\sqrt{\epsilon_n}}{1+\sqrt{\epsilon_n}}\right)
\max_{x\in(x_{k},x_{k+1})} \left[\frac{\Delta
x_k}{x-\xi\lambda_n(1-\rho_n(1-x))}\right]-1 \nonumber \\
\end{eqnarray}
Hence
\begin{eqnarray}
T_k & \geq &
\left(\frac{1-\sqrt{\epsilon_n}}{1+\sqrt{\epsilon_n}}\right)
\left[\frac{\Delta
x_k}{x_k-\xi\lambda_n(1-\rho_n(1-x_k))}\right]-1
\nonumber \\
\end{eqnarray}
Note that
\begin{eqnarray}
& & \frac{\Delta x_k}{x_k-\xi\lambda_n(1-\rho_n(1-x_k))} \nonumber \\
& & = \min\left\{ \frac{1}{\sqrt{\epsilon_n}},
\frac{\xi/2}{x_k-\xi\lambda_n(1-\rho_n(1-x_k))}
\right\}\rightarrow \infty \mbox{ as } n\rightarrow \infty
\nonumber \\
\end{eqnarray}
From the above, we conclude that the claim is true.

We bound the ratio $m/T_\eta$ as follows.
\begin{eqnarray}
\frac{m}{T_\eta} \leq \left(\frac{m}{m-1}\right)
\frac{m-1}{\sum_{k=0}^{m-2}T_k} \nonumber \\
\end{eqnarray}
From the above claim, we have
\begin{eqnarray}
\frac{m-1}{\sum_{k=0}^{m-2}T_k} \rightarrow 0 \mbox{ as }n
\rightarrow \infty
\end{eqnarray}
We conclude that the ratio $m/T_\eta$ goes to $0$ as $n$ goes to
infinity. The theorem is proven.
\end{proof}

\section{The Proof of Lemma
\ref{lemma_basic_degree_distribution1}}
\label{proof_lemma_basic_degree_distribution1}

\begin{proof}
We can check that $\hat{\gamma}(x)$ is a valid degree
distribution, $\sum_i\hat{\gamma}_i=1$ and the average degree is
$a$, $\int_{0}^{1}\hat{\gamma}(x)=1/a$.

Note that $\frac{i}{i+2}$ and $1$ are two roots of the polynomial
\begin{equation}
x-\frac{i+2}{2(i+1)}x^2-\frac{i}{2(i+1)}=\frac{\hat{\gamma}(x)-\gamma(x)}{x^{i-1}}
\end{equation}
Hence, for $0<x<\frac{i}{i+2}$,
\begin{equation}
\hat{\gamma}(x)<\gamma(x)
\end{equation}
for $\frac{i}{i+2}<x<1$,
\begin{equation}
\hat{\gamma}(x)>\gamma(x)
\end{equation}
\end{proof}

\comment{

\section{The Proof of Theorem
\ref{theorem_gamma_min}} \label{proof_theorem_gamma_min}

\begin{proof}
Let us assume that the optimal $\gamma(x)$ has a non-zero
$\gamma_i$, $2<i<d$. Since $x<(1/2)$, $x<(i-1)/(i+1)$, for all
$i\in \{3,\cdots,d-1\}$. Then another degree distribution
\begin{align}
\hat{\gamma}(x)= & { \gamma(x) - \beta x^{i-1}
+\left\{\frac{i-1}{2i}\beta
x^{i-2}+\frac{i+1}{2i}\beta x^{i}\right\} } \notag \\
\end{align}
can be constructed, where $\beta$ is a sufficiently small positive
real number. Similar as the discussion in Lemma
\ref{lemma_basic_degree_distribution1}, we can show that
$\hat{\gamma}(x)>\gamma(x)$. This results in a contradiction.
Hence the theorem is proven.
\end{proof}

}

\section{The Proof of Theorem
\ref{theorem_gamma_max}} \label{proof_theorem_gamma_max}

\begin{proof}

We prove the theorem by contradiction. Assume that the degree
distribution $\gamma(x)$ is nonzero for more than three indices or
 is nonzero for two non-consecutive indices. Then, either one of
the following two cases happens.

\emph{Case 1:} there exist three consecutive indices $i-1$, $i$,
$i+1$ such that $\gamma_{i-1}$, $\gamma_i$, and $\gamma_{i+1}$ are
nonzero.

Note that
\begin{equation}
x>1-\frac{2}{d+2}>\frac{i-1}{i+1}
\end{equation}
According to Lemma \ref{lemma_basic_degree_distribution1}, we can
constructed another degree distribution $\hat{\gamma}(x)$ such
that $\hat{\gamma}(x)>\gamma(x)$. This contradict to the
hypothesis that $\gamma(x)$ is optimal.

\emph{Case 2:} there exist positive integers $i$ and $j$ such that
$\gamma_{i}$, $\gamma_{j}$ are nonzero, $i<i+1<j$, and
$\gamma_{k}=0$ for any $k$, $i<k<j$.

The conditions of Lemma \ref{lemma_basic_degree_distribution1} is
also satisfied in this case. Define $\gamma^k(x)$ for each $k$,
$i<k<j$, as:
\begin{equation}
\gamma^k(x)=x^{k-1} - \frac{k}{2(k+1)}\beta
x^{k-2}-\frac{k+2}{2(k+1)} x^{k}
\end{equation}
We can find real numbers
$\alpha_{i+1},\alpha_{i+2},\cdots,\alpha_{j-1}$ such that
$\alpha_k>0$, for $i<k<j$, and the polynomial
\begin{equation}
\hat{\gamma}(x)=\gamma (x)+\sum_{k=i+1}^{j-1}\alpha_k \gamma^k(x)
\end{equation}
is a valid degree distribution. The degree distribution
$\hat{\gamma}(x)$ has average right degree $a$. Since
$\gamma^k(x)>0$ for each $k$, we have
\begin{equation}
\hat{\gamma}(x)>\gamma(x)
\end{equation}
This contradict to the hypothesis that $\gamma(x)$ is optimal.

The theorem follows from the discussions in the two cases.
\end{proof}

\section{The Proof of Theorem
\ref{right_concen_main_thm}} \label{proof_right_concen_main_thm}

\begin{proof}

The optimal $\rho_j^\ast$ must also be the solution of the
following constrain optimization problem with $\lambda_i$ being
fixed and equal to $\lambda_i^\ast$.
\begin{equation}
\min \int_{\eta}^{\xi}\frac{1}{x-\xi\lambda(1-\rho(1-x))}dx
\end{equation}
subject to
\begin{equation}
\xi\lambda^\ast(1-\rho^\ast(1-x))<x,\,\,\,\mbox{ for any
}x\in(0,\xi]
\end{equation}
\begin{equation}
\frac{\int_0^1\rho(x)dx}{\int_0^1\lambda(x)dx}= 1-C+\Delta R
\end{equation}
\begin{equation}
\sum_j \rho_j=1
\end{equation}

According to the Karush-Kuhn-Tucker condition \cite{Chong01},
$\rho^\ast_j$ satisfy the following equation, for all $j$ with
nonzero $\rho^\ast_j$,
\begin{equation}
\frac{\partial}{\partial \rho_j} F(\lambda,\rho,\eta)+ \alpha
\frac{\partial}{\partial
\rho_j}\left[\frac{\int_0^1\rho(x)dx}{\int_0^1\lambda(x)dx}\right]+
\beta\left[\frac{\partial}{\partial \rho_j}\sum_j \rho_j\right]=0
\end{equation}
where $\alpha$, $\beta$ are constants. This is equivalent to
\begin{equation}
j\frac{\partial}{\partial \rho_j} F(\lambda,\rho,\eta)+
\frac{\alpha}{\int_0^1\lambda(x)dx}+\beta j=0
\end{equation}
After finding the derivative of $F(\lambda,\rho,\eta)$, we can
rewrite the above equation as follows:
\begin{equation}
\int_{\eta}^{\xi} g(x) j(1-x)^{j-1} dx=
\alpha\frac{1}{\int_0^1\lambda(x)dx}+\beta j
\label{condition_optimal}
\end{equation}
where
\begin{equation}
g(x)= \frac {\xi \sum_i \left\{\lambda_i (i-1)
\left[1-\sum_j\rho_j(1-x)^{j-1}\right]^{i-2}\right\}}
{\left\{x-\xi\sum_i\left[\lambda_i\left(1-\sum_j\rho_j(1-x)^{j-1}\right)^{i-1}
\right]\right\}^2}
\end{equation}

In Case 1, $j(1-x)^{j-1}$ is a concave function with respect to
$j$. Hence $\int_{\eta}^{\xi} g(x) j(1-x)^{j-1} dx$ is also
concave with respect to $j$. There exist at most two $j$'s which
satisfy Eqn. \ref{condition_optimal}. The theorem is proven in
this case.

In Case 2, it suffices to show that if $\min_{x\in
(1-e^{-2/d_c},\xi)}[x-\xi\lambda^\ast(1-\rho^\ast(1-x))]$ is
bounded from below, then $\rho_j^\ast$ are nonzero only for two
indices when $\eta$ is sufficiently small.

Note that
\begin{align}
& { \frac{\partial^2}{\partial j^2}\left[\int_{\eta}^{\xi} g(x)
j(1-x)^{j-1} dx \right]= } \notag \\
& { \frac{\partial^2}{\partial
j^2}\left[\int_{\eta}^{1-\exp(-2/d_c)} g(x) j(1-x)^{j-1} dx
\right]+ \frac{\partial^2}{\partial
j^2}\left[\int_{1-\exp(-2/d_c)}^{\xi} g(x) j(1-x)^{j-1} dx \right]
} \label{two_part_formula}
\end{align}
We will show that the first term at the right hand side of Eqn.
\ref{two_part_formula} goes to $-\infty$ as $\eta$ goes to zero
and the second term is bounded. When $\eta$ is sufficiently small,
$\frac{\partial^2}{\partial j^2}\left[\int_{\eta}^{\xi} g(x)
j(1-x)^{j-1} dx \right]$ is negative for $2\leq j \leq d_c$. Hence
$\int_{\eta}^{\xi} g(x) j(1-x)^{j-1} dx$ is concave with respect
to $j$. There exist at most two $j$'s which satisfy Eqn.
(\ref{condition_optimal}). The theorem is proven.

We first show the bounding of the second term. Assume for $x\in
(1-e^{-2/d_c},\xi)$
\begin{equation}
[x-\xi\lambda^\ast(1-\rho^\ast(1-x))]\geq {\cal M}
\end{equation}
For $x\in(1-e^{-2/d_c},\xi)$, we can bound $g(x)$ as follows:
\begin{equation}
g(x)=\frac {\xi \sum_i \left[\lambda_i (i-1)
\left(1-\sum_j\rho_j(1-x)^{j-1}\right)^{i-2}\right]}
{\left\{x-\xi\sum_i\lambda_i\left[1-\sum_j\rho_j(1-x)^{j-1}\right]^{i-1}\right\}^2}
\leq \frac {\xi d_c}{{\cal M}^2}
\end{equation}
Hence, we have the following bound:
\begin{equation}
\left|\frac{\partial^2}{\partial
j^2}\left[\int_{1-\exp(-2/d_c)}^{\xi} g(x) j(1-x)^{j-1} dx
\right]\right|\leq \frac {\xi d_c}{{\cal M}^2}
\left|\int_{1-\exp(-2/d_c)}^{\xi} \frac{\partial^2}{\partial
j^2}\left[j(1-x)^{j-1}\right] dx \right|
\end{equation}
We can further bound $\frac{\partial^2}{\partial j^2}
\left[j(1-x)^{j-1}\right]$ as follows:
\begin{align}
\frac{\partial^2}{\partial j^2} \left[j(1-x)^{j-1}\right] & { =
\left|(1-x)^{j-1}\log(1-x)[2+j\log(1-x)]\right|} \notag \\
& { \leq \left |\log(1-\xi)\right|(2+d_c\left|\log(1-\xi)\right|)
}
\end{align}
Hence, we have the following bound:
\begin{equation}
\left|\frac{\partial^2}{\partial^2
j}\left[\int_{1-\exp(-2/d_c)}^{\xi} g(x) j(1-x)^{j-1} dx
\right]\right|\leq \frac {\xi d_c}{{\cal M}^2}
(\xi-1-\exp(-2/d_c)) \left
|\log(1-\xi)\right|(2+d_c\left|\log(1-\xi)\right|
\end{equation}

Next, we will show that the first term in the right hand side of
Eqn. \ref{two_part_formula} goes to $-\infty$ as $\eta$ goes to
zero.

According to theorem \ref{theorem_flatness_main}, as $x\rightarrow
0$,
\begin{equation}
\left\{x-\xi\sum_i\lambda_i\left[1-\sum_j\rho_j(1-x)^{j-1}\right]^{i-1}\right\}^2
\leq O(x^2)
\end{equation}
\begin{equation}
\xi \sum_i \lambda_i (i-1)
\left[1-\sum_j\rho_j(1-x)^{j-1}\right]^{i-1}=O(x)
\end{equation}
We also have
\begin{equation}
\frac{\partial^2}{\partial j^2}\left[j(1-x)^{j-1}\right]  =
(1-x)^{j-1}\log(1-x)\left[2+j\log(1-x)\right]= O(x)
\end{equation}
\begin{equation}
[1-\sum_j\rho_j(1-x)^{j-1}] = O(x)
\end{equation}
Therefore
\begin{align}
 g(x)\left\{\frac{\partial^2}{\partial j^2}\left[j(1-x)^{j-1}\right]\right\} = &
 {\frac {\left\{\xi \sum_i \lambda_i (i-1)
\left[1-\sum_j\rho_j(1-x)^{j-1}\right]^{i-2}\right\}\frac{\partial
^2}{\partial j^2}\left[j(1-x)^{j-1}\right]}
{\left\{x-\xi\sum_i\lambda_i\left[1-\sum_j\rho_j(1-x)^{j-1}\right]^{i-1}\right\}^2}}
\notag \\
  \geq &
 {\frac{O(x)O(x)}{O(x^2)\left[1-\sum_j\rho(1-x)^{j-1}\right]}
 =  \frac{1}{O(x)}}
\end{align}
This implies that
\begin{equation}
\frac{\partial^2}{\partial j^2}\left[\int_{\eta}^{1-\exp(-2/d_c)}
g(x) j(1-x)^{j-1} dx \right]\rightarrow \infty
\end{equation}
as $\eta\rightarrow 0$. When $\eta$ is sufficiently small,
$\int_{\eta}^{\xi} g(x) j(1-x)^{j-1} dx$ is concave with respect
to $j$. There exist at most two $j$'s which satisfy Eqn.
(\ref{condition_optimal}). The theorem is proven.
\end{proof}

\end{document}